\documentclass[12pt] {article}
\usepackage{psfrag}
\usepackage{graphicx}
\usepackage{latexsym,amsfonts}
\usepackage{amssymb}
\begin{document}
\setcounter{page}{1}
\def\theequation{\arabic{section}.\arabic{equation}}
\def\theequation{\thesection.\arabic{equation}}
\setcounter{section}{0}

\title{On spontaneous breaking of continuous symmetry in
1+1--dimensional space--time}

\author{M. Faber\thanks{E--mail: faber@kph.tuwien.ac.at, Tel.:
+43--1--58801--14261, Fax: +43--1--5864203} ~~and~
A. N. Ivanov\thanks{E--mail: ivanov@kph.tuwien.ac.at, Tel.:
+43--1--58801--14261, Fax: +43--1--5864203}~\thanks{Permanent Address:
State Polytechnical University, Department of Nuclear Physics, 195251
St. Petersburg, Russian Federation}}

\date{\today}

\maketitle
\vspace{-0.5in}
\begin{center}
{\it Atominstitut der \"Osterreichischen Universit\"aten,
Arbeitsbereich Kernphysik und Nukleare Astrophysik, Technische
Universit\"at Wien, \\ Wiedner Hauptstr. 8-10, A-1040 Wien,
\"Osterreich }
\end{center}

\begin{center}
\begin{abstract}
We analyse Coleman's theorem asserting the absence of Goldstone bosons
and spontaneously broken continuous symmetry in the quantum field
theory of a free massless (pseudo)scalar field in 1+1--dimensional
space--time (Comm. Math. Phys. {\bf 31}, 259 (1973)). We confirm that
Coleman's theorem reproduces the well--known statement by Wightman
about the non--existence of a quantum field theory of a free massless
(pseudo)scalar field in 1+1--dimensional space--time in terms of
Wightman's observables defined on the test functions from ${\cal
S}(\mathbb{R}^{\,2})$.  Referring to our results (Eur. Phys. J. C {\bf
24}, 653 (2002)) we argue that a formulation of a quantum field theory
of a free massless (pseudo)scalar field in terms of Wightman's
observables defined on the test functions from ${\cal
S}_0(\mathbb{R}^{\,2})$ is motivated well by the possibility to remove
the collective zero--mode of the ``center of mass'' motion of a free
massless (pseudo)scalar field (Eur. Phys. J. C {\bf 24}, 653 (2002))
responsible for infrared divergences of the Wightman functions. We
show that in the quantum field theory of a free massless
(pseudo)scalar field with Wightman's observables defined on the test
functions from ${\cal S}_0(\mathbb{R}^{\,2})$ the continuous symmetry
can be spontaneously broken.  Coleman's theorem reformulated for the
test functions from ${\cal S}_0(\mathbb{R}^{\,2})$ does not refute
this result. We construct the most general version of a quantum field
theory of a self--coupled massless (pseudo)scalar field with a
conserved current. We show that this theory satisfies Wightman's
axioms and {\it Wightman's positive definiteness condition} with
Wightman's observables defined on the test functions from ${\cal
S}(\mathbb{R}^{\,2})$ and possesses a spontaneously broken continuous
symmetry. Nevertheless, in this theory the generating functional of
Green functions exists only when the collective zero--mode is not
excited by the external source.
\end{abstract}
\end{center}

\newpage

\section{Introduction}
\setcounter{equation}{0}

\hspace{0.2in} In the literature the absence of a spontaneously broken
continuous symmetry \cite{MWH} and Goldstone bosons \cite{SC73} in
quantum field theories in two dimensional space--time is related
\cite{IZ80}--\cite{KH98} to the Mermin--Wagner--Hohenberg theorem
\cite{MWH}, asserting the vanishing of spontaneous {\it magnetization}
or {\it long--range order} in spin systems (Heisenberg models, the XY
model and so on) at non--zero temperature, and Coleman's proof of the
non--existence of Goldstone bosons in the quantum field theory of a
free massless (pseudo)scalar field \cite{SC73}. Since we are
interested in quantum field theories at zero--temperature we analyse
below Coleman's theorem \cite{SC73} only.

The absence of Goldstone bosons has been related by Coleman
\cite{SC73} to the problem of the infrared divergences of the Wightman
functions of a free massless (pseudo)scalar field which we call
$\vartheta(x)$
\begin{eqnarray}\label{label1.1}
\hspace{-0.3in}&&D^{(+)}(x; \mu) = \langle
\Psi_0|\vartheta(x)\vartheta(0)|\Psi_0\rangle = \int
\frac{d^2k}{(2\pi)^2}\,F^{(+)}(k)\,e^{\textstyle -\,i\,k\cdot x}
=\nonumber\\ \hspace{-0.3in}&&= \int \frac{d^2k}{2\pi}\,\theta(+
k^0)\,\delta(k^2)\,e^{\textstyle -\,i\,k\cdot x}
=\frac{1}{2\pi}\int^{\infty}_{-\infty}\frac{dk^1}{2k^0}\,e^{\textstyle
-\,i\,k\cdot x} = - \frac{1}{4\pi}\,{\ell n}[-\mu^2x^2 +
i\,0\cdot\varepsilon(x^0)], \nonumber\\ \hspace{-0.3in}&&D^{(-)}(x;
\mu) = \langle \Psi_0|\vartheta(0)\vartheta(x)|\Psi_0\rangle = \int
\frac{d^2k}{(2\pi)^2}\,F^{(-)}(k)\,e^{\textstyle +\,i\,k\cdot x}
=\nonumber\\ \hspace{-0.3in}&&= \int \frac{d^2k}{(2\pi)^2}\,\theta(-
k^0)\,\delta(k^2)\,e^{\textstyle +\,i\,k\cdot x} =
\frac{1}{2\pi}\int^{\infty}_{-\infty}\frac{dk^1}{2k^0}\,e^{\textstyle
+\,i\,k\cdot x} = - \frac{1}{4\pi}\,{\ell n}[-\mu^2x^2 -
i\,0\cdot\varepsilon(x^0)],\nonumber\\ \hspace{-0.3in}&&
\end{eqnarray}
where $F^{(\pm)}(k) = 2\pi\,\theta(\pm k^0)\,\delta(k^2)$ are the
Fourier transforms of the Wightman functions, $\theta(k^0)$ is the
Heaviside function, $\delta(k^2)$ is the Dirac $\delta$--function of
$k^2 = (k^0)^2 - (k^1)^2$ in 2--dimensional momentum space,
$\varepsilon(x^0)$ is the sign function, $x^2 = (x^0)^2 - (x^1)^2$,
$k\cdot x = k^0x^0 - k^1x^1$, $k^0 = |k^1|$ is the energy of free
massless (pseudo)scalar quantum with momentum $k^1$ and $\mu$ is the
infrared cut--off reflecting the infrared divergences of the Wightman
functions (\ref{label1.1}). As has been stated by Klaiber in his
seminal paper \cite{BK67}: {\it If one wants to solve the Thirring
model, one has to overcome this problem.}\,\footnote{In this
connection we would like to refer to the solution of the massless
Thirring model suggested by Hagen \cite{CH67}, who has succeeded in
solving the massless Thirring model avoiding the problem of infrared
divergences of free massless (pseudo)scalar boson fields. The problem
of infrared divergences does not appear within the path--integral
solution of the massless Thirring model as well \cite{KF}.}

Recently \cite{FI1} we have shown that the fermion fields in the
massless Thirring model evolve via a phase of spontaneously broken
chiral $U(1)\times U(1)$ symmetry. The wave function of the ground
state of the massless Thirring model in the chirally broken phase
coincides with the wave function of the superconducting phase of the
Bardeen--Cooper--Schrieffer (BCS) theory of superconductivity.  For
the quantum field theory of a free massless (pseudo)scalar field we
have shown \cite{FI1}--\cite{FI12} that this theory satisfies all
requirements for a continuous symmetry to be spontaneously broken: (i)
the ground state is not invariant under the continuous symmetry
\cite{FI1}--\cite{FI12}, (ii) the energy level of the ground state is
infinitely degenerate \cite{FI1,FI8, FI9} and (iii) Goldstone bosons
appear \cite{FI1,FI2,FI10} and they are the quanta of a free massless
(pseudo)scalar field.

In this paper we would like to show that the results obtained in
\cite{FI1}--\cite{FI12} do not contradict Coleman's theorem
\cite{SC73} and Wightman's axioms \cite{AW64}--\cite{JG81}.

The quantum field theory of a free massless (pseudo)scalar field
$\vartheta(x)$ is described by the Lagrangian \cite{SC73,FI1,FI2}
\begin{eqnarray}\label{label1.2}
{\cal L}(x) =
\frac{1}{2}\,\partial_{\mu}\vartheta(x)\partial^{\mu}\vartheta(x),
\end{eqnarray}
where $x = (x^0,x^1)$ is a 2--vector, which is invariant under field
translations \cite{SC73,FI1,FI2}
\begin{eqnarray}\label{label1.3}
 \vartheta(x) \to \vartheta\,'(x) = \vartheta(x) + \alpha,
\end{eqnarray}
where $\alpha$ is an arbitrary parameter $\alpha \in \mathbb{R}^1$
\cite{FI1,FI2}\,\footnote{As has been shown in \cite{FI1} the
parameter $\alpha = - 2\alpha_{\rm A}$ is related to the chiral phase
$\alpha_{\rm A}$ of global chiral rotations of Thirring fermion
fields.}.

The conserved current associated with these field translations is
equal to
\begin{eqnarray}\label{label1.4}
j_{\mu}(x) = \partial_{\mu}\vartheta(x).
\end{eqnarray}
The total {\it charge} is defined by the time--component of
$j_{\mu}(x)$ \cite{FI1,FI2}
\begin{eqnarray}\label{label1.5}
Q(x^0) =\lim_{L\to \infty}Q_L(x^0) = \lim_{L\to \infty} \int^{ L/2}_{-
L/2}dx^1\,j_0(x^0,x^1) = \lim_{L\to \infty} \int^{ L/2}_{-
L/2}dx^1\,\frac{\partial}{\partial x^0}\vartheta(x^0,x^1),
\end{eqnarray}
where $L$ is the volume of the system. The time--component
$j_0(x^0,x^1)$ of the current $j_{\mu}(x)$ coincides with the
conjugate momentum of the $\vartheta$--field, $j_0(x^0,x^1) =
\dot{\vartheta}(x^0,x^1) = \Pi(x^0,x^1)$. Due to this the operators
$j_0(x^0,x^1)$ and $\vartheta(y^0,y^1)$ obey the equal--time
commutation relation
\begin{eqnarray}\label{label1.6}
[j_0(x^0,x^1),\vartheta(x^0,y^1)] = [\Pi(x^0,x^1),\vartheta(x^0,y^1)]
= -i\,\delta(x^1 - y^1).
\end{eqnarray}
This is the canonical commutation relation which leads to
\begin{eqnarray}\label{label1.7}
i\,[Q(x^0),\vartheta(x)] = 1.
\end{eqnarray}
As a result the total {\it charge} operator $Q(x^0)$ given by
(\ref{label1.5}) generates shifts of the $\vartheta$--field
\begin{eqnarray}\label{label1.8}
\vartheta\,'(x) &=& e^{\textstyle +i\alpha
Q(x^0)}\,\vartheta(x)\,e^{\textstyle - i\alpha Q(x^0)} = \vartheta(x)
+ (i\alpha)\,[Q(x^0),\vartheta(x)] \nonumber\\&& +
\frac{1}{2!}(i\alpha)^2\,[Q(x^0),[Q(x^0),\vartheta(x)]] + \ldots =
\vartheta(x) + \alpha.
\end{eqnarray}
For the further analysis it is convenient to denote
\begin{eqnarray}\label{label1.9}
\delta \vartheta(x) = \alpha \,i\,[Q(x^0),\vartheta(x)].
\end{eqnarray}
According to Goldstone's theorem \cite{JG61,KY76} the criterion
for the existence of Goldstone bosons is the non--vanishing vacuum
expectation value of $\delta \vartheta(x)$. This reads
\begin{eqnarray}\label{label1.10}
\langle \Psi_0|\delta \vartheta(x)|\Psi_0 \rangle = \lim_{L \to
\infty}\alpha\,i\langle \Psi_0|[Q_L(x^0),\vartheta(x)]|\Psi_0 \rangle
\neq 0,
\end{eqnarray}
where $|\Psi_0\rangle$ is a vacuum wave function.  

For the subsequent analysis of a free massless (pseudo)scalar field
theory we use the expansion of the massless (pseudo)scalar field
$\vartheta(x)$ and the conjugate momentum $\Pi(x)$ into plane waves
\begin{eqnarray}\label{label1.11}
\vartheta(x) &=&
\int^{\infty}_{-\infty}\frac{dk^1}{2\pi}\,\frac{1}{2k^0}\,
\Big(a(k^1)\,e^{\textstyle -i\,k\cdot x} +
a^{\dagger}(k^1)\,e^{\textstyle i\,k\cdot x}\Big),\nonumber\\ \Pi(x)
&=& \int^{\infty}_{-\infty}\frac{dk^1}{2\pi}\,\frac{1}{2i}\,
\Big(a(k^1)\,e^{\textstyle -i\,k\cdot x} -
a^{\dagger}(k^1)\,e^{\textstyle i\,k\cdot x}\Big),
\end{eqnarray}
where $a(k^1)$ and $a^{\dagger}(k^1)$ are annihilation and creation
operators obeying the standard commutation relation
\begin{eqnarray}\label{label1.12}
[a(k^1), a^{\dagger}(q^1)] = (2\pi)\,2k^0\,\delta(k^1 - q^1).
\end{eqnarray}
This gives the canonical commutation relation (\ref{label1.6}).

According to Coleman's proof \cite{SC73} the vacuum expectation value
$\langle \Psi_0|\delta \vartheta(x)|\Psi_0 \rangle$ should vanish,
i.e.
\begin{eqnarray}\label{label1.13}
\langle \Psi_0|\delta \vartheta(x)|\Psi_0 \rangle = 0.
\end{eqnarray}
This has been interpreted \cite{SC73}--\cite{KH98} as the absence of
Goldstone bosons and the proof of the impossibility for the continuous
symmetry to be spontaneously broken in 1+1--dimensional space--time.

It is well--known that the spontaneous breaking of a continuous
symmetry occurs when the ground state of the system is not invariant
under the symmetry group \cite{IZ80,JG61}. As has been shown in
\cite{FI1,FI2,FI8} the ground state of the system described by the
Lagrangian (\ref{label1.2}) is not invariant under field translations
(\ref{label1.3}). Therefore, the field--translation symmetry should be
spontaneously broken and Goldstone bosons should appear
\cite{FI1}--\cite{FI10}. According to Witten's criterion for Goldstone
bosons \cite{EW78}, Goldstone bosons should saturate low--energy
theorems and Ward identities. As we have shown in \cite{FI10}, the
quanta of the free massless (pseudo)scalar field $\vartheta(x)$,
describing the bosonized massless Thirring model with fermion fields
quantized in the chirally broken phase, saturate the low--energy
theorems and axial--vector Ward identities. This means that they
satisfy Witten's criterion for Goldstone bosons \cite{EW78}. The
non--invariance of the ground state of the free massless
(pseudo)scalar field $\vartheta(x)$, described by the Lagrangian
(\ref{label1.2}), can be demonstrated by acting with the operator
$e^{\textstyle +i\alpha Q(0)}$ on the wave function of the ground
state.

Recently \cite{FI8,FI9} we have shown that the BCS--type wave function
of the ground state of the massless Thirring model in the chirally
broken phase \cite{FI1} bosonizes to the form
\begin{eqnarray}\label{label1.14}
|\Omega_{\,0}\rangle = \exp\Big\{i\,\frac{\pi}{2}\,\frac{M}{g}
\int^{\infty}_{-\infty}dx^1\,\sin\beta\vartheta(0,x^1)\Big\}
|\Psi_0\rangle,
\end{eqnarray}
where $M$ is a dynamical mass of Thirring fermion fields in the
chirally broken phase, $g$ is the coupling constant of the Thirring
model \cite{FI1}.

Under the symmetry transformation induced by the operator
$e^{\textstyle +i\alpha Q(0)}$ the wave function (\ref{label1.14})
transforms as follows \cite{FI8,FI9}
\begin{eqnarray}\label{label1.15}
|\Omega_{\,\alpha}\rangle = e^{\textstyle +i\alpha
Q(0)}|\Omega_{\,0}\rangle = \exp\Big\{i\,\frac{\pi}{2}\,\frac{M}{g}
\int^{\infty}_{-\infty}dx^1\,\sin\beta(\vartheta(0,x^1) -
\alpha)\Big\}|\Psi_0\rangle.
\end{eqnarray}
The wave functions $|\Omega_{\,\alpha}\rangle$ and
$|\Omega_{\,\alpha\,'}\rangle$ are orthogonal and normalized to unity
\cite{FI8,FI9}
\begin{eqnarray}\label{label1.16}
\langle \Omega_{\,\alpha\,'}|\Omega_{\,\alpha}\rangle =
\delta_{\alpha\,'\alpha}.
\end{eqnarray}
In \cite{FI2,FI8} we have shown that within the Schwinger formulation
of the quantum field theory \cite{JS69} the amplitude of the {\it
vacuum--to--vacuum} transition $\langle
\Omega^+_{\,0}|\Omega^-_{\,0}\rangle_J$, where $J(x)$ is an external
source of a free massless (pseudo)scalar field $\vartheta(x)$ and
$|\Omega^-_{\,0}\rangle$ and $\langle \Omega^+_0|$ are the vacuum
states at $T = - \infty$ and $T = + \infty$, coincides with the
generating functional of Green functions $Z[J]$ for the free massless
(pseudo)scalar field $\vartheta(x)$, described by the Lagrangian
(\ref{label1.2}), defined by
\begin{eqnarray}\label{label1.17}
Z[J] = \int {\cal D}\vartheta\,\exp\Big\{i\int
d^2x\,\Big[\frac{1}{2}\partial_{\mu}\vartheta(x)
\partial^{\mu}\vartheta(x) + \vartheta(x)J(x)]\Big\}.
\end{eqnarray}
In order to get a non--vanishing value for $Z[J]$ we have to impose
the constraint of the external source \cite{FI2}
\begin{eqnarray}\label{label1.18}
\int d^2x\,J(x) = 0,
\end{eqnarray}
which provides the suppression of the excitation of the collective
zero--mode of the free massless (pseudo)scalar field $\vartheta(x)$ by
the external source $J(x)$. 

The collective zero--mode is responsible for the ``center of mass''
motion \cite{FI2} and is the origin of the infrared divergences of the
two--point Wightman functions in the quantum field theory of the free
massless (pseudo)scalar field \cite{FI9}. Hence, a removal of the
``center of mass'' motion allows to describe correlation functions in
such a quantum field theory, defined by the generating functional of
Green functions $Z[J]$ with the constraint (\ref{label1.18}), only in
terms of the vibrational modes \cite{FI2,FI8,FI9}.

The time--dependent wave function (\ref{label1.14}) can be found in
the usual way using the translation formula \cite{FI8}
\begin{eqnarray}\label{label1.19}
|\Omega^T_{\,0} \rangle = e^{\textstyle\,+ iH T}|\Omega_{\,0} \rangle
&=& e^{\textstyle\,+ iH T}\exp\Big\{i\,\frac{\pi}{2}\,\frac{M}{g}
\int^{\infty}_{-\infty}dx^1\,\sin\beta\vartheta(0,x^1)\Big\}
e^{\textstyle\,-iH T}|\Psi_0\rangle =\nonumber\\ &=&
\exp\Big\{i\,\frac{\pi}{2}\,\frac{M}{g}
\int^{\infty}_{-\infty}dx^1\,\sin\beta\vartheta(T,x^1)\Big\}
|\Psi_0\rangle,
\end{eqnarray}
where $H$ is the Hamilton operator of the free massless (pseudo)scalar
field \cite{FI2}
\begin{eqnarray}\label{label1.20}
H = \frac{1}{2}\int^{\infty}_{-\infty}dx^1\,\Bigg[\Big(\frac{\partial
\vartheta(x)}{\partial x^0}\Big)^2 + \Big(\frac{\partial
\vartheta(x)}{\partial x^1}\Big)^2\Bigg].
\end{eqnarray}
Then, we have taken into account that $H|\Psi_0\rangle = 0$
\cite{FI2,FI9}.

The paper is organized as follows. In Section 2 we discuss a possible
physical interpretation of the test functions for Wightman's
observables. We argue that in the case of a quantum field theory of
the free massless (pseudo)scalar field $\vartheta(x)$, described by
the Lagrangian (\ref{label1.2}), Wightman's observables should be
defined on the test functions from ${\cal S}_0(\mathbb{R}^{\,2})$.
Such a reduction of the class of the test functions corresponds to the
immeasurably of the collective zero--mode of the free massless
(pseudo)scalar field describing the motion of the ``center of
mass''. As has been shown in \cite{FI2} the collective zero--mode is
irrelevant to the dynamics of the free massless (pseudo)scalar field.
The reduction of the test functions from ${\cal S}(\mathbb{R}^{\,2})$
to ${\cal S}_0(\mathbb{R}^{\,2})$ reconciles the problem of the
correct formulation of a quantum field theory of a free massless
(pseudo)scalar field within the path--integral approach in terms of
the generating functional of Green functions (or the Schwinger
external source approach) with Wightman's axiomatic quantum field
theory in terms of Wightman's observables \cite{FI7}.  In Section 3 we
show that the spontaneous breaking of continuous symmetry in the
quantum field theory of a free massless (pseudo)scalar field with
Wightman's observables defined on the test functions from ${\cal
S}_0(\mathbb{R}^{\,2})$ does not contradict Coleman's theorem
\cite{SC73}.  In Section 4 we construct a quantum field theory of a
self--coupled massless (pseudo)scalar field with a conserved current
using the K\"allen--Lehmann representation for the description of
two--point correlation functions. We show that Wightman's observables
in such a theory can be defined on the test functions from ${\cal
S}(\mathbb{R}^{\,2})$. However, a non--vanishing value of the
generating functional of Green functions can be gained only when the
constraint on the external source (\ref{label1.18}) is fulfilled. This
agrees fully with Hasenfratz's analysis of non--linear
two--dimensional $\sigma$--models with $O(N)$ symmetry \cite{PH84}.
In the Conclusion we discuss the obtained results.

\section{Wightman's axioms and quantum field theory of a free 
massless (pseudo)scalar field with spontaneously broken continuous
symmetry} 
\setcounter{equation}{0}

\hspace{0.2in} A quantum field theory in 1+1--dimensional space--time
is well--defined within the axiomatic approach
\cite{AW64,AW80} if it satisfies the following set of
Wightman's axioms \cite{AW80,JG81}:
\begin{itemize}
\item {\bf W1} (Covariance). There is a continuous unitary
representation of the imhomogeneous Lorentz group $g \to U(g)$ on the
Hilbert space ${\cal H}$ of quantum theory states. The generators $H =
(P^0, P^1)$ of the translation subgroup have spectrum in the forward
cone $(p^0)^2 - (p^1)^2 \ge 0,\,p^0 \ge 0$. There is a vector
$|\Psi_0\rangle \in {\cal H}$ (the vacuum) invariant under the
operators $U(g)$.

\item {\bf W2} (Observables). There are field operators
$\{\vartheta(h): h(x) \in {\cal S}(\mathbb{R}^{\,2})\}$ densely
defined on ${\cal H}$. The vector $|\Psi_0\rangle$ is in the domain of
any polynomial in the $\vartheta(h)$'s, and the subspace ${\cal H}\,'$
spanned algebraically by the vectors $\{\vartheta(h_1)\ldots
\vartheta(h_n)|\Psi_0\rangle; n \ge 0, h_i \in {\cal
S}(\mathbb{R}^{\,2})\}$ is dense in ${\cal H}$. The field
$\vartheta(h)$ is covariant under the action of the Lorentz group on
${\cal H}$, and depends linearly on $h$. In particular,
$U^{\dagger}(g) \vartheta(h)U(g) = \vartheta(h_g)$.

\item {\bf W3} (Locality). If the supports of $h(x)$ and
$h\,'(x)$ are space--like separated, then
$[\vartheta(h),\vartheta(h\,')] = 0$ on ${\cal H}\,'$.

\item {\bf W4} (Vacuum). The vacuum vector $|\Psi_0\rangle$ is the
unique vector (up to scalar multiples) in ${\cal H}$ which is
invariant under time translations.
\end{itemize}

These axioms should be supplemented by {\it Wightman's positive
definiteness condition} \cite{AW80} related to the positivity of the
norm of a quantum state in the quantum field theory under
consideration. Indeed, for quantum states $|\Psi\rangle$ defined by
\begin{eqnarray}\label{label2.1}
|\Psi\rangle &=& \alpha_0|\Psi_0\rangle + \alpha_1\int d^2x_1 h(x_1)
\vartheta(x_1)|\Psi_0\rangle \nonumber\\
\hspace{-0.5in}&+&  \frac{\alpha_2}{2!}\int\!\!\!\int
d^2x_1d^2x_2 h(x_1) h(x_2)\vartheta(x_1) \vartheta(x_2)|\Psi_0\rangle
+ \ldots,
\end{eqnarray}
which are superpositions of $n$--particle quantum states
$|\Psi_n\rangle$
\begin{eqnarray}\label{label2.2}
|\Psi_n\rangle = \frac{1}{\sqrt{n!}}\int\!\ldots\!\int d^2x_1\ldots
d^2x_n\,h(x_1)\ldots h(x_n)\,\vartheta(x_1)\ldots
\vartheta(x_n)\,|\Psi_0\rangle,
\end{eqnarray}
the vectors in the Hilbert space ${\cal H}$
\cite{AW64,AW80}, should have a positive norm
\begin{eqnarray}\label{label2.3}
&&\|\Psi\| = \Big\| \alpha_0|\Psi_0\rangle + \alpha_1\int
d^2x_1\,h(x_1)\,\vartheta(x_1)|\Psi_0\rangle\nonumber\\ &&\hspace{1in}
+ \frac{\alpha_2}{2!}\int\!\!\!\int d^2x_1d^2x_2\,h(x_1)
h(x_2)\,\vartheta(x_1)\,\vartheta(x_2)|\Psi_0\rangle + \ldots
\Big\|\ge 0
\end{eqnarray}
for all $\alpha_i\in \mathbb{R}^1\,(i=0,1,\ldots)$ and test
functions $h(x)$ from the Schwartz class ${\cal S}(\mathbb{R}^{\,2})$
\cite{AW64}.  In a quantum field theory of a free field the
inequality (\ref{label2.3}) reduces to the constraint \cite{AW80}
\begin{eqnarray}\label{label2.4}
\int\!\!\!\int d^2xd^2y\,h^*(x)\,D^{(+)}(x-y; \mu)\,h(y) \ge 0,
\end{eqnarray}
which is the {\it Wightman positive definiteness condition}.

The problem of the correct formulation of a quantum field theory of a
free massless (pseudo)scalar field $\vartheta(x)$ in agreement with
Wightman's axioms and {\it Wightman's positive definiteness condition}
spans many years and has a long history\cite{AW64,BS63}. The main
problem concerns the observation that the two--point Wightman function
$D^{(+)}(x;\mu)$ (\ref{label1.1}) does not satisfy {\it Wightman's
positive definiteness condition} on the test functions $h(x)$ from the
Schwartz class ${\cal S}(\mathbb{R}^{\,2})$ \cite{AW64}.

As has been pointed out by Wightman \cite{AW64} due to infrared
divergences of the two--point Wightman function $D^{(+)}(x-y; \mu)$
one cannot formulate a quantum field theory of a free massless
(pseudo)scalar field $\vartheta(x)$ on the class of test functions
from ${\cal S}(\mathbb{R}^{\,2})$ consistent with {\it Wightman's
positive definiteness condition}. Therefore, in the sense of the
non--existence of a quantum state $|\Psi\rangle$ defined by
(\ref{label2.1}) with a positive norm, the quantum field theory of a
free massless (pseudo)scalar field $\vartheta(x)$ with Wightman's
observables defined on the test functions $h(x)$ from the Schwartz
class ${\cal S}(\mathbb{R}^{\,2})$ does not exist \cite{AW64}.

In order to avoid the problem of the violation of {\it Wightman's
positive definiteness condition} in a quantum field theory of a free
massless (pseudo)scalar field Wightman has noticed that one can define
Wightman's observables on the test functions $h(x)$ from the Schwartz
class ${\cal S}_0(\mathbb{R}^{\,2}) = \{h(x) \in {\cal
S}(\mathbb{R}^{\,2});\tilde{h}(0) = 0\} \subset {\cal
S}(\mathbb{R}^{\,2})$ instead of ${\cal S}(\mathbb{R}^{\,2})$.

As has been shown in \cite{FI2} the non--existence of a quantum field
theory of a free massless (pseudo)scalar field in terms of the
generating functional of Green functions $Z[J]$ or the amplitude of
the {\it vacuum--vacuum} transitions $\langle
\Omega^+_0|\Omega^-_0\rangle_J$ is related to the contribution of the
collective zero--mode of the $\vartheta$--field describing the motion
of the ``center of mass''. Such a collective zero--mode can be removed
from the system without influence on the evolution of vibrational
modes \cite{FI2}. A {\it necessary} and {\it sufficient} condition for
the removal of the collective zero--mode of the $\vartheta$--field is
the constraint (\ref{label1.18}). 

In order to reconcile Schwinger's formulation of a quantum field
theory of a free massless (pseudo)scalar field in the form suggested
in \cite{FI2} and Wightman's axiomatic approach one has to understand
how the constraint (\ref{label1.18}) is related to the choice of the
Schwartz class of test functions $h(x)$. 

For the analysis of this relation we suggest to interpret the test
functions $h(x)$ as the {\it apparatus functions} of the detector,
which an observer uses for measurements of quanta of the
$\vartheta$--field in terms of matrix elements of Wightman's
observables $\vartheta(h)$ defined by
\begin{eqnarray}\label{label2.5}
\vartheta(h) = \int d^2x\,h(x)\,\vartheta(x).
\end{eqnarray}
For example, in terms of $\vartheta(h)$ the matrix element $\langle
\Psi_0|\vartheta(h)|k^1\rangle$, where $|k^1\rangle =
a^{\dagger}(k^1)|\Psi_0\rangle$ is a one--particle state with momentum
$k^1$, should describe the amplitude of the registration of a massless
quantum of the $\vartheta$--field with momentum $k^1$ by the
detector. The quantity $P_{\rm det}(k^1) = |\langle \Psi_0|
\vartheta(h) |k^1 \rangle|^2 $ has the meaning of the probability to
detect a massless quantum of the $\vartheta$--field with momentum
$k^1$.

In terms of the Fourier transform $\tilde{h}(k^0,k^1)$ of the test
function $h(x)$ the probability $P_{\rm det}(k^1) = |\langle
\Psi_0|\vartheta(h)|k^1\rangle|^2$ is equal to
\begin{eqnarray}\label{label2.6}
P_{\rm det}(k^1) = |\tilde{h}(k^0,k^1)|^2 \neq 1.
\end{eqnarray}
The probability $P_{\rm free}(k^1)$ of a massless quantum of the
$\vartheta$--field with momentum $k^1$ to be out the detector is equal
to
\begin{eqnarray}\label{label2.7}
P_{\rm free}(k^1) = 1 - P_{\rm det}(k^1) = 1 -
|\tilde{h}(k^0,k^1)|^2.
\end{eqnarray}
Hence, if the test functions are equal to zero, $h(x) = \tilde{h}(k) =
0$, this should mean that there are no devices which can detect
massless quanta of the $\vartheta$--field, a massless quantum with a
momentum $k^1$ should be free with a probability $P_{\rm free}(k^1) =
1$.

This assertion can be fully confirmed by a direct calculation of
$P_{\rm free}(k^1)$ in terms of the matrix element $\langle
\Psi_0|\vartheta(x)|k^1\rangle$, where the field $\vartheta(x)$ is
defined by the plane wave expansion (\ref{label1.11}). One obtains
\begin{eqnarray}\label{label2.8}
P_{\rm free}(k^1) = |\langle \Psi_0|\vartheta(x)|k^1\rangle|^2 = 1.
\end{eqnarray}
Therefore, the function $\tilde{h}(k^0, k^1)$ should be treated as a
characteristic of the detector, the {\it apparatus function} related to
the {\it resolving power} of the device. 

One can notice that the probability $P_{\rm det}(k^1)$, defined by
(\ref{label2.6}), is a regular function of $k^1$ in the limit $k^1 \to
0$. Taking the limit $k^1 \to 0$ in (\ref{label2.6}) we get
\begin{eqnarray}\label{label2.9}
\lim_{k^1 \to 0}P_{\rm det}(k^1) = |\tilde{h}(0,0)|^2.
\end{eqnarray}
Thus, the quantity $|\tilde{h}(0,0)|^2$ describes the probability of
the detection of the infrared quanta $(k^0 = k^1 = 0)$. A more
explicit meaning of the quantity $|\tilde{h}(0,0)|^2$ can be derived
from the definition of Wightman's observable (\ref{label2.5}). Indeed,
as has been stated in the Introduction the shift of the field
$\vartheta(x) \to \vartheta\,'(x) = \vartheta(x) + \alpha$ corresponds
to the shift of the ``center of mass'' of the
$\vartheta$--field. Under the shift $\vartheta(x) \to \vartheta\,'(x)
= \vartheta(x) + \alpha$ Wightman's observable (\ref{label2.5})
changes as follows
\begin{eqnarray}\label{label2.10}
\vartheta(h) \to \vartheta\,'(h) &=& \int d^2x\,[\vartheta(x) +
\alpha]\,h(x) = \int d^2x\,\vartheta(x)\,h(x) + \alpha\int d^2x\,h(x)
=\nonumber\\
&=&\int d^2x\,\vartheta(x)\,h(x) + \alpha\,\tilde{h}(0,0).
\end{eqnarray}
It is seen that the test function $\tilde{h}(0,0)$ feels the motion of
the ``center of mass'' of the free massless (pseudo)scalar field
$\vartheta(x)$ defined by the collective zero--mode. Since the
collective zero--mode can deleted from the states defining correlation
functions in terms of the generating functional of Green functions
$Z[J]$, one can use the detectors which are insensitive to the
collective zero--mode. This can be gained by setting
\begin{eqnarray}\label{label2.11}
\tilde{h}(0,0) = 0.
\end{eqnarray}
Therefore, in our interpretation the constraint $\tilde{h}(0,0) = 0$
means that the detector is insensitive to the collective zero--mode of
the field $\vartheta(x)$. Therefore, the requirement to define a
quantum field theory of a free massless (pseudo)scalar field on the
test functions from ${\cal S}_0(\mathbb{R}^{\,2}) = \{h(x) \in {\cal
S}(\mathbb{R}^{\,2}); \tilde{h}(0,0) = 0\}$ would correspond to the
exclusion of the collective zero--mode from the observable states of
the quantum field $\vartheta(x)$ in terms of Wightman's observables
$\vartheta(h)$.

Now let us show that a quantum field theory of a free massless
(pseudo)scalar field $\vartheta(x)$, described by the Lagrangian
(\ref{label1.2}), satisfies all Wightman's axioms and {\it Wightman's
positive definiteness condition} on the test functions from ${\cal
S}_0(\mathbb{R}^{\,2})$ and unstable under spontaneous breaking of
continuous symmetry (\ref{label1.3}).

As the validity of Wightman's axioms {\bf W1} (Covariance) and {\bf
W2} (Observables) is obviously fulfilled on the class of test
functions $h(x) \in {\cal S}_0(\mathbb{R}^{\,2})$, and according to
{\bf W4} (Vacuum) the vacuum state is invariant under time
translations, let us verify the fulfillment of Wightman's axiom {\bf
W3} (Locality). For this aim we have to analyse the commutator
\begin{eqnarray}\label{label2.12}
[\vartheta(h),\vartheta(h\,'\,)] = \int\!\!\!\int
d^2x\,d^2y\,h(x)\,h\,'(y)\,[\vartheta(x),\vartheta(y)].
\end{eqnarray}
Since the commutator $[\vartheta(x),\vartheta(y)]$ is equal to
\begin{eqnarray}\label{label2.13}
[\vartheta(x),\vartheta(y)] = D^{(+)}(x-y; \mu) - D^{(-)}(x-y; \mu) =
- \frac{i}{2}\,\varepsilon(x^0 - y^0)\,\theta((x-y)^2),
\end{eqnarray}
where $\theta((x-y)^2)$ is the Heaviside function, the r.h.s. of
(\ref{label2.12}) reads
\begin{eqnarray}\label{label2.14}
[\vartheta(h),\vartheta(h\,'\,)] = \frac{i}{2}\int\!\!\!\int
d^2x\,d^2y\,h(x)\,h\,'(y)\,\varepsilon(x^0 - y^0)\,\theta((x-y)^2).
\end{eqnarray}
Due to the presence of the Heaviside function $\theta((x-y)^2)$ it is
obvious that the integrand vanishes if the supports of $h(x)$ and
$h\,'(y)$ are space--like separated, i.e. $(x-y)^2 < 0$. It follows
\begin{eqnarray}\label{label2.15}
[\vartheta(h),\vartheta(h\,'\,)] = 0.
\end{eqnarray}
This corroborates the validity of Wightman's axiom {\bf W3} (Locality)
within the quantum field theory of a free massless (pseudo)scalar
field $\vartheta(x)$ defined on the Schwartz class of test functions
${\cal S}_0(\mathbb{R}^{\,2}) = \{h(x) \in {\cal S}(\mathbb{R}^{\,2});
\tilde{h}(0,0) = 0\}$.

Now we have to verify the fulfillment of {\it Wightman's positive
definiteness condition} (\ref{label2.4}). Using the Fourier transform
$F^{(+)}(k) = 2\pi\,\theta(k^0)\,\delta(k^2)$ of the Wightman function
$D^{(+)}(x-y;\mu)$ (\ref{label1.1}) and passing to the light--cone
variables $k_+ = k^0 + k^1$, $k_- = k^0 - k^1$ and $d^2k =
\frac{1}{2}dk_+dk_-$ we get
\begin{eqnarray}\label{label2.16}
&&\int\!\!\!\int d^2x d^2y\,h^*(x)\,D^{(+)}(x-y; \mu)\,h(y) =\int
\frac{d^2k}{2\pi}\,|\tilde{h}(k)|^2\theta(k^0)\,\delta(k^2)
=\nonumber\\ &&= \int^{\infty}_{-\infty}\int^{\infty}_{-\infty}
\frac{dk_+dk_-}{4\pi}\,|\tilde{h}(k_+,k_-)|^2
\Bigg[\,\frac{\theta(k_+)}{k_+}\,\delta(k_-) +
\frac{\theta(k_-)}{k_-}\,\delta(k_+)\,\Bigg] =\nonumber\\ &&=
\frac{1}{2\pi} \int^{\infty}_0\frac{dk_+}{k_+}\,|\tilde{h}(k_+,0)|^2.
\end{eqnarray}
Since the test functions $h(x)$ belong to ${\cal
S}_0(\mathbb{R}^{\,2}) = \{h(x) \in {\cal S}(\mathbb{R}^{\,2});
\tilde{h}(0,0) = 0\}$, the integral over $k_+$ is positive defined and
convergent.

Thus, a quantum field theory of a free massless (pseudo)scalar field
$\vartheta(x)$ is well--defined on the class of test functions $h(x)$
belonging to ${\cal S}_0(\mathbb{R}^{\,2}) = \{h(x) \in {\cal
S}(\mathbb{R}^{\,2}); \tilde{h}(0,0) = 0\}$. All Wightman's axioms
including {\it Wightman's positive definiteness condition} are
fulfilled. The physical reason of the formulation of the quantum field
theory of a free massless (pseudo)scalar field $\vartheta(x)$ on the
test functions from ${\cal S}_0(\mathbb{R}^{\,2})$ is the
insignificance of the collective zero--mode for the evolution of
vibrational modes \cite{FI2}.

As we have shown in \cite{FI2} such a quantum field theory of a free
massless (pseudo)scalar field is unstable under spontaneous breaking
of continuous symmetry (\ref{label1.3}). We would like to confirm this
statement by calculating the spontaneous {\it magnetization}.

For this aim we suggest to consider the massless (pseudo)scalar field
$\vartheta(x)$ coupled to an external ``magnetic'' field
$h_{\lambda}(x)$ \cite{ID2}, where $h_{\lambda}(x)$ is a sequence of
the Schwartz functions from ${\cal S}_0(\mathbb{R}^{\,2})$ with
vanishing norm at $\lambda \to \infty$. The Lagrangian
(\ref{label1.2}) should be changed as follows
\begin{eqnarray}\label{label2.17}
{\cal L}(x; h_{\lambda}) =
\frac{1}{2}\,\partial_{\mu}\vartheta(x)\partial^{\mu}\vartheta(x) +
h_{\lambda}(x)\,\vartheta(x).
\end{eqnarray}
The Lagrangian (\ref{label2.17}) defines the action of a massless
(pseudo)scalar field $\vartheta(x)$ coupled to the ``magnetic''field
$h_{\lambda}(x)$
\begin{eqnarray}\label{label2.18}
S[\vartheta, h_{\lambda}] = \int d^2x\,{\cal L}(x; h_{\lambda}) =
\frac{1}{2}\int
d^2x\,\partial_{\mu}\vartheta(x)\partial^{\mu}\vartheta(x) +\int d^2x\,
h_{\lambda}(x)\,\vartheta(x).
\end{eqnarray}
Since the ``magnetic'' field $h_{\lambda}(x)$ belongs to the Schwartz
class ${\cal S}_0(\mathbb{R}^{\,2})$ obeying the constraint
\begin{eqnarray}\label{label2.19}
\int d^2x\,h_{\lambda}(x) = \tilde{h}_{\lambda}(0) = 0,
\end{eqnarray}
the action $S[\vartheta, h_{\lambda}]$ is invariant under the symmetry
transformation (\ref{label1.2}).

By the field--shift (\ref{label1.2}) we get
\begin{eqnarray}\label{label2.20}
\hspace{-0.3in}&&S[\vartheta, h_{\lambda}] \to S\,'[\vartheta,
h_{\lambda}] = \frac{1}{2}\int
d^2x\,\partial_{\mu}\vartheta\,'(x)\partial^{\mu}\vartheta\,'(x) +\int
d^2x\, h_{\lambda}(x)\,\vartheta\,'(x) =\nonumber\\ \hspace{-0.3in}&&=
S[\vartheta, h_{\lambda}] + \alpha\int d^2x\,h_{\lambda}(x).
\end{eqnarray}
Due to the constraint (\ref{label2.19}) the r.h.s. of (\ref{label2.20})
is equal to $S[\vartheta, h_{\lambda}]$. This confirms the invariance
of the action under the symmetry transformations (\ref{label1.2}).

According to Itzykson and Drouffe \cite{ID2} the magnetization ${\cal
M}(h_{\lambda})$ can be defined by \cite{FI2}
\begin{eqnarray}\label{label2.21}
\hspace{-0.3in}&&{\cal M}(h_{\lambda}) = \langle
\Psi_0|\cos\vartheta(h_{\lambda})|\Psi_0\rangle = \lim_{\mu\to 0}\exp
\Big\{ - \frac{1}{2}\int d^2x\,d^2y\,h^*_{\lambda}(x)\,D^{(+)}(x-y;
\mu)\,h_{\lambda}(y)\Big\} = \nonumber\\ \hspace{-0.3in}&&= \exp
\Big\{ - \frac{1}{4\pi}\int^{\infty}_0\frac{dk_+}{k_+}\,
|\tilde{h}_{\lambda}(k_+, 0)|^2\,\Big\}.
\end{eqnarray}
The momentum integral in the exponent of the r.h.s. of
(\ref{label2.21}) is convergent. If we switch off the ``magnetic''
field taking the limit $h_{\lambda} \to 0$, this can be carried out
adiabatically defining $h_{\lambda}(x) = e^{\textstyle
-\varepsilon\,\lambda}\,h(x)$ for $\lambda \to \infty$ with
$\varepsilon$, a positive, infinitesimally small parameter, we get
\begin{eqnarray}\label{label2.22}
{\cal M} = \lim_{\lambda \to \infty}{\cal M}(h_{\lambda}) = 1.
\end{eqnarray}
This agrees with our results obtained in \cite{FI2}. The quantity
${\cal M}$ is the spontaneous {\it magnetization}. Since the
spontaneous magnetization does not vanish, ${\cal M} = 1$, the
continuous symmetry, caused by the field--shifts (\ref{label1.2}), is
spontaneously broken. This confirms our statement concerning the
existence of the chirally broken phase in the massless Thirring model
\cite{FI1}.

Now let us analyse the properties of the field operator $\vartheta(h)$
under the symmetry transformations (\ref{label1.3}). The result
obtained in (\ref{label2.12}) can be also proved by acting with the
total {\it charge} operator $Q(x^0)$ defined by (\ref{label1.5}) or
(\ref{label1.16}). We get
\begin{eqnarray}\label{label2.23}
\hspace{-0.5in}&&\vartheta\,'(h) = e^{\textstyle +i\alpha
Q(x^0)}\,\vartheta(h)\,e^{\textstyle - i\alpha Q(x^0)} = \int
d^2y\,h(y)\, e^{\textstyle +i\alpha
Q(x^0)}\,\vartheta(y)\,e^{\textstyle - i\alpha Q(x^0)} =
\vartheta(h)\nonumber\\
\hspace{-0.5in}&&+ i\alpha\int d^2y\,h(y)\,[Q(x^0),\vartheta(y)] =
\vartheta(h) + i\alpha\int
d^2y\,h(y)\int^{\infty}_{-\infty}dx^1\frac{\partial }{\partial
x^0}\,[\vartheta(x),\vartheta(y)].
\end{eqnarray}
Since the commutator $[\vartheta(x),\vartheta(y)]$ is defined by
(\ref{label2.13}), the time derivative of this commutator reads
\begin{eqnarray}\label{label2.24}
\frac{\partial }{\partial x^0}\,[\vartheta(x),\vartheta(y)] = -
i\,|x^0 - y^0|\,\delta((x-y)^2) = \frac{1}{2i}\,\delta(x_+ - y_+) +
\frac{1}{2i}\,\delta(x_- - y_-).
\end{eqnarray}
Substituting (\ref{label2.24}) in (\ref{label2.23}), integrating over
$x^1$ and using the constraint (\ref{label2.6}) we obtain
\begin{eqnarray}\label{label2.25}
\vartheta\,'(h) = e^{\textstyle +i\alpha
Q(x^0)}\,\vartheta(h)\,e^{\textstyle - i\alpha Q(x^0)} = \vartheta(h)
+ \alpha\int d^2y\,h(y) = \vartheta(h).
\end{eqnarray}
Thus, we have found that Wightman's observables $\vartheta(h)$ are
invariant under $\vartheta$--field shifts. In this sense the field
operator $\vartheta(h)$ can be really treated as an observable to the
same extent as electric $\vec{E}(t,\vec{r}\,)$ and magnetic
$\vec{B}(t,\vec{r}\,)$ fields, invariant under gauge transformations
and measurable, whereas the vector potential $A^{\mu}(t,\vec{r}\,) =
(\varphi(t,\vec{r}\,),\vec{A}(t,\vec{r}\,))$, which is not invariant
under gauge transformations and, correspondingly, an immeasurable
quantity.

However, the invariance of the field operator $\vartheta(h)$ under 
symmetry transformations (\ref{label1.3}) tells nothing about the
non--existence of Goldstone bosons in the quantum field theory of the
free massless (pseudo)scalar field $\vartheta(x)$ described by the
Lagrangian (\ref{label1.2}).

Indeed, the invariance of the field operator $\vartheta(h)$ provides
the vanishing of the field variation $\delta \vartheta(h) = 0$ in the
strong sense at the operator level. It is not related to the peculiar
property of the vacuum wave function $|\Psi_0\rangle$ to be invariant
under the symmetry transformations (\ref{label1.3}). Wightman's axiom
{\bf W4} (Vacuum), demanding the invariance of the vacuum wave
function $|\Psi_0\rangle$ under time translations, does not require
the invariance of $|\Psi_0\rangle$ under any internal symmetry group
like that inducing the field shifts (\ref{label1.3}).

Hence, Wightman's quantum field theory of a free massless
(pseudo)scalar field $\vartheta(x)$, formulated on the test functions
$h(x)$ from ${\cal S}_0(\mathbb{R}^{\,2}) = \{h(x) \in {\cal
S}(\mathbb{R}^{\,2}); \tilde{h}(0,0) = 0\}$, gives a nice possibility
to deal with well--defined observable quantities but does not clarify
the problem of the absence of Goldstone bosons and spontaneously
broken symmetry in 1+1--dimensional space--time. 

\section{Coleman's theorem} 
\setcounter{equation}{0}

\hspace{0.2in} Coleman's theorem \cite{SC73}, asserting the absence of
Goldstone bosons and spontaneously broken continuous symmetry in
1+1--dimensional quantum field theories, is closely related to
Wightman`s statement about the non--existence of a 1+1--dimensional
quantum field theory of a free massless (pseudo)scalar field with
Wightman's observables defined on the test functions from ${\cal
S}(\mathbb{R}^{\,2})$ \cite{AW64}.

The problem of the absence of Goldstone bosons and spontaneously
broken continuous symmetry in 1+1--dimensional quantum field theories
Coleman has investigated in terms of the Fourier transforms of the
two--point correlation functions
\begin{eqnarray}\label{label3.1}
F^{(+)}(k) &=&\int d^2x\,e^{\textstyle i\,k\cdot x}\,\langle
\Psi_0|\vartheta(x)\vartheta(0)|\Psi_0\rangle,\nonumber\\
F^{(+)}_{\mu}(k) &=&i\int d^2x\,e^{\textstyle i\,k\cdot x}\,\langle
\Psi_0|j_{\mu}(x)\vartheta(0)|\Psi_0\rangle,\nonumber\\
F^{(+)}_{\mu\nu}(k) &=&\int d^2x\,e^{\textstyle i\,k\cdot x}\,\langle
\Psi_0|j_{\mu}(x)j_{\nu}(0)|\Psi_0\rangle,
\end{eqnarray}
where $F^{(+)}(k)$ is the Fourier transform of the Wightman function.
The Fourier transform $F^{(+)}_{\mu}(k)$ Coleman has determined by the
expression \cite{SC73}
\begin{eqnarray}\label{label3.2}
F^{(+)}_{\mu}(k) = \sigma\,k_{\mu}\theta(k^0)\delta(k^2),
\end{eqnarray}
where $\sigma$ is a parameter. Using the expression (\ref{label1.10})
the parameter $\sigma$ can be related to the vacuum expectation value
$\langle \Psi_0|\delta \vartheta(0)|\Psi_0\rangle$. For this aim we
suggest to consider the relation
\begin{eqnarray}\label{label3.3}
&&\int^{\infty}_{-\infty}dk^0\,F^{(+)}_0(k^0,0)
=i\,\frac{1}{2}\int^{\infty}_{-\infty}dx^1
\int^{\infty}_{\infty}dx^0\int^{\infty}_{-\infty}dk^0\,e^{\textstyle
\,i\,k^0x^0}\langle
\Psi_0|[j_0(x^0,x^1),\vartheta(0)]|\Psi_0\rangle.\nonumber\\ &&
\end{eqnarray}
The r.h.s. of (\ref{label3.3}) is obtained due to the fact that
$F^{(+)}_{\mu}(k^0,k^1)$ is a real function of $k^0$ and $k^1$. Using
the expression (\ref{label3.4}) and integrating over $k^0$ and $x^0$
we get
\begin{eqnarray}\label{label3.4}
\frac{1}{2}\,\sigma =i\,\pi\,\int^{\infty}_{-\infty}dx^1\,\langle
\Psi_0|[j_0(0,x^1),\vartheta(0)]|\Psi_0\rangle.
\end{eqnarray}
In terms of the total {\it charge} $Q(0)$ related to the current
$j_0(0,x^1)$ by (\ref{label1.5}) we can transcribe (\ref{label3.4})
into the form
\begin{eqnarray}\label{label3.5}
i\langle \Psi_0|[Q(0),\vartheta(0)]|\Psi_0\rangle = \frac{\sigma}{2\pi}.
\end{eqnarray}
Due to equation (\ref{label1.10}) the vacuum expectation value
$\langle \Psi_0|\delta \vartheta(0)|\Psi_0\rangle$ can be expressed in
terms of the parameter $\sigma$ and reads
\begin{eqnarray}\label{label3.6}
\langle \Psi_0|\delta \vartheta(0)|\Psi_0\rangle =\alpha\, 
\frac{\sigma}{2\pi}.
\end{eqnarray}
If $\sigma = 0$ this gives $\langle \Psi_0|\delta
\vartheta(0)|\Psi_0\rangle = 0$ and according to the Goldstone theorem
\cite{JG61} this should testify the absence of Goldstone bosons and
spontaneously broken continuous symmetry. 

For the proof of his theorem Coleman has considered the
Cauchy--Schwarz inequality
\begin{eqnarray}\label{label3.7}
\int \frac{d^2k}{2\pi}\,|\tilde{h}_{\lambda}(k)|^2\,F^{(+)}(k)\int
\frac{d^2k}{2\pi}\,|\tilde{h}_{\lambda}(k)|^2\,F^{(+)}_{00}(k)\ge
\Big[\int
\frac{d^2k}{2\pi}\,|\tilde{h}_{\lambda}(k)|^2\,F^{(+)}_0(k)\Big]^2,
\end{eqnarray}
where $\tilde{h}_{\lambda}(k)$ is the Fourier
transform 
\begin{eqnarray}\label{label3.8}
\tilde{h}_{\lambda}(k) = \tilde{f}(\lambda k_-)\,\tilde{g}(k_+) +
\tilde{f}(\lambda k_+)\,\tilde{g}(k_-)
\end{eqnarray}
of the sequence of test function $h_{\lambda}(x)$ at $\lambda \to
\infty$ defined by
\begin{eqnarray}\label{label3.9}
h_{\lambda}(x) = h_{\lambda}(x_+,x_-) =
\frac{1}{\lambda}\,f\Big(\frac{x_+}{\lambda}\Big)\,g(x_-) +
\frac{1}{\lambda}\,f\Big(\frac{x_-}{\lambda}\Big)\,g(x_+),
\end{eqnarray}
where $k_- = k^0 - k^1, k_+ = k^0 + k^1$ and $x_+ = (x^0 - x^1)/2, x_-
= (x^0 - x^1)/2$ are the light--cone variables in momentum space and
coordinate space--time.

According to Coleman the test functions (\ref{label3.9}) should belong
to the Schwartz class ${\cal S}(\mathbb{R}^{\,1})\otimes {\cal
S}_0(\mathbb{R}^{\,1})$, where $f(x_{\pm}) \in {\cal
S}(\mathbb{R}^{\,1})$ with $\tilde{f}(0) \neq 0$ and $g(x_{\pm}) \in
{\cal S}_0(\mathbb{R}^{\,1}) = \{g(x_{\pm}) \in {\cal
S}(\mathbb{R}^{\,1}); \tilde{g}(0) = 0\}$ \cite{AW64}.

For the analysis of the Cauchy--Schwarz inequality (\ref{label3.7})
Coleman has formulated a lemma
\begin{eqnarray}\label{label3.10}
\lim_{\lambda \to
\infty}\int^{\infty}_{-\infty}dk_-\,\tilde{f}(\lambda k_-)\,F(k_+,k_-)
= c\,\delta(k_+),
\end{eqnarray}
where $c$ is a positive constant and $F(k_+,k_-)$ is a positive
Lorentz--invariant distribution \cite{SC73}.

Due to this lemma and the requirement for the test functions
$f(x_{\pm})$ to belong to the Schwarz class ${\cal
S}(\mathbb{R}^{\,1})$ the massless mode with $k^2 = k_+k_- = 0$ is
excluded from the intermediate states defining the Fourier transform
of the Wightman function $F^{(+)}(k)$. Indeed, the massless mode
contribution to $F^{(+)}(k)$ should be proportional to
$\theta(k^0)\delta(k^2) = \theta(k_+ + k_-)\delta(k_+k_-)$ (see
(\ref{label1.1})). The contribution of this term to (\ref{label3.10})
is given by
\begin{eqnarray}\label{label3.11}
\hspace{-0.3in}&&
\int^{\infty}_{-\infty}dk_-\,\tilde{f}(\lambda
k_-)\,\theta(k_+ + k_-)\,\delta(k_-k_+) = 
\nonumber\\
\hspace{-0.3in}&&= \int^{\infty}_{-\infty}dk_-\, \tilde{f}(\lambda
k_-)\, \theta(k_+ + k_-)\,\Bigg[\frac{1}{|k_-|}\,\delta(k_+) +
\frac{1}{|k_+|}\,\delta(k_-) \Bigg]=\nonumber\\ \hspace{-0.3in}&& =
\delta(k_+)\int^{\infty}_{-\infty}dk_-\,\tilde{f}(\lambda
k_-)\,\frac{\theta(k_-)}{|k_-|} +
\tilde{f}(0)\,\frac{\theta(k_+)}{k_+} =\nonumber\\
\hspace{-0.3in}&&=\delta(k_+)\int^{\infty}_0\frac{dk_-}{k_-}
\,\tilde{f}(\lambda k_-) + \tilde{f}(0)\,\frac{\theta(k_+)}{k_+} =
c(\lambda)\,\delta(k_+) + \tilde{f}(0)\,\frac{\theta(k_+)}{k_+}.
\end{eqnarray}
The constant $c(\lambda)$ is defined by the integral
\begin{eqnarray}\label{label3.12}
c(\lambda) = \int^{\infty}_0\frac{dk_-}{k_-}\, \tilde{f}(\lambda k_-)
= \int^{\infty}_0\frac{dk_-}{k_-}\,\tilde{f}(k_-),
\end{eqnarray}
where we have made a change of variables $\lambda\,k_- \to k_-$. The
r.h.s. of (\ref{label3.11}) satisfies Coleman's lemma
(\ref{label3.10}) if and only if
\begin{eqnarray}\label{label3.13}
\tilde{f}(0) = 0.
\end{eqnarray}
However, the constraint (\ref{label3.13}) is fulfilled only for the
test functions from ${\cal S}_0(\mathbb{R}^{\,1})$. If the test
functions $f(x_{\pm})$ belong to ${\cal S}(\mathbb{R}^{\,1})$ with
$\tilde{f}(0) \neq 0$, as it has been assumed by Coleman, the
coefficient $c(\lambda)$ given by (\ref{label3.12}) is divergent in
the infrared region. This makes the relation (\ref{label3.11})
meaningless. Hence, due to Coleman's lemma (\ref{label3.10}) the
function $\theta(k^0)\delta(k^2) = \theta(k_+ + k_-)\delta(k_+k_-)$ is
not a well--defined {\it tempered} distribution \cite{AW64}.

Thus, if $\vartheta(x)$ is a free massless (pseudo)scalar field,
described by the Lagrangian (\ref{label1.1}) with the Fourier
transform $F^{(+)}(k) \propto \theta(k^0)\delta(k^2)$ of the
two--point Wightman function, the existence of this field is
prohibited by Coleman's lemma (\ref{label3.10}). This agrees with
Wightman's statement if Wightman's observables are defined on the test
functions from ${\cal S}(\mathbb{R}^{\,2})$.

However, as we have shown in Section 2 in the case of a quantum field
theory of a free massless (pseudo)scalar field Wightman's observables
should be defined on the test functions from ${\cal
S}_0(\mathbb{R}^{\,1})$. In agreement with this reduction Coleman's
requirement $f(x_{\pm}) \in {\cal S}(\mathbb{R}^{\,1})$ can be
weakened and replaced by $f(x_{\pm}) \in {\cal S}_0(\mathbb{R}^{\,1})$
with $\tilde{f}(0) = 0$. As it is shown above on the test functions
$f(x_{\pm}) \in {\cal S}_0(\mathbb{R}^{\,1})$ the Fourier transform
$F^{(+)}(k) = 2\pi\,\theta(k^0)\,\delta(k^2)$ of the Wightman function
$D^{(+)}(x; \mu)$ (\ref{label1.1}) is well--defined {\it tempered}
distribution \cite{AW64}.

Let us show that Coleman's constraint $\sigma = 0$ is not a solution
of the Cauchy--Schwarz inequality (\ref{label3.7}) on the test
functions (\ref{label3.9}) but the consequence of the lemma
(\ref{label3.10}). The Fourier transforms (\ref{label3.1}) defined for
a free massless (pseudo)scalar field $\vartheta(x)$ with $j_{\mu}(x) =
\partial_{\mu}\vartheta(x)$ are equal to
\begin{eqnarray}\label{label3.14}
F^{(+)}(k) &=&\sigma\,\theta(k^0)\delta(k^2),\nonumber\\
F^{(+)}_{\mu}(k)
&=&\sigma\,k_{\mu}\,\theta(k^0)\delta(k^2),\nonumber\\
F^{(+)}_{\mu\nu}(k) &=&\sigma\,k_{\mu}k_{\nu}\,\theta(k^0)\delta(k^2),
\end{eqnarray}
where the canonical value of the parameter $\sigma$ is $\sigma =
2\pi$. Nevertheless, following Coleman we keep it arbitrary and try to
find the constraint from the solution of the Cauchy--Schwarz
inequality defined on the test functions (\ref{label3.8}).

For convenience of the further analysis we suggest to rewrite the
Cauchy--Schwarz inequality (\ref{label3.7}) as follows 
\begin{eqnarray}\label{label3.15}
J(\lambda)\,J_{00}(\lambda)\ge J^2_0(\lambda),
\end{eqnarray}
where $J(\lambda)$, $J_0(\lambda)$ and $J_{00}(\lambda)$ are momentum
integrals of $F^{(+)}(k)$, $F^{(+)}_0(k)$ and $F^{(+)}_{00}(k)$
multiplied by $|\tilde{h}_{\lambda}(k)|^2/2\pi$, respectively.
\begin{eqnarray}\label{label3.16}
\hspace{-0.3in}&&J(\lambda) = \int
\frac{d^2k}{2\pi}\,|\tilde{h}_{\lambda}(k)|^2\,F^{(+)}(k) =\nonumber\\
\hspace{-0.3in}&&=\frac{\sigma}{4\pi}
\int^{\infty}_{-\infty}dk_-\int^{\infty}_{-\infty}dk_+\,
|\tilde{f}(\lambda
k_-)\,\tilde{g}(k_+) + \tilde{f}(\lambda k_+)\,\tilde{g}(k_-)|^2\,
\theta(k_+ +
k_-)\,\delta(k_-k_+)=\nonumber\\
\hspace{-0.3in}&&= \frac{\sigma}{4\pi}
\int^{\infty}_{-\infty}dk_-\,|\tilde{f}(\lambda k_-)\,\tilde{g}(0) +
\tilde{f}(0)\,\tilde{g}(k_-)|^2\, \frac{\theta(k_-)}{k_-}\nonumber\\
\hspace{-0.3in}&&+ \frac{\sigma}{4\pi}
\int^{\infty}_{-\infty}dk_+\,|\tilde{f}(\lambda k_+)\, \tilde{g}(0) +
\tilde{f}(0)\,\tilde{g}(k_+)|^2\, \frac{\theta(k_+)}{k_+} =\nonumber\\
\hspace{-0.3in}&&= \frac{\sigma}{2\pi}
\int^{\infty}_0\frac{dk_+}{k_+}\,|\tilde{f}(\lambda k_+)\,
\tilde{g}(0) + \tilde{f}(0)\,\tilde{g}(k_+)|^2=\nonumber\\
\hspace{-0.3in}&&=\frac{\sigma}{2\pi}\,|\tilde{g}(0)|^2
\int^{\infty}_0\frac{dk_+}{k_+}\,|\tilde{f}(\lambda k_+)|^2 + 
\frac{\sigma}{2\pi}\,\tilde{f}(0)\tilde{g}^*(0)
\int^{\infty}_0\frac{dk_+}{k_+}\,
\tilde{f}^*(\lambda k_+)\tilde{g}(k_+)\nonumber\\
\hspace{-0.3in}&&+ \frac{\sigma}{2\pi}\,\tilde{f}^*(0)
\tilde{g}(0)\int^{\infty}_0 \frac{dk_+}{k_+}\,\tilde{f}(\lambda
k_+)\tilde{g}^*(k_+) + \frac{\sigma}{2\pi}\, |\tilde{f}(0)|^2
\int^{\infty}_0\frac{dk_+}{k_+}\,|\tilde{g}(k_+)|^2 = \nonumber\\ &&
=\frac{\sigma}{2\pi}\, |\tilde{f}(0)|^2
\int^{\infty}_0\frac{dk_+}{k_+}\,|\tilde{g}(k_+)|^2,
\end{eqnarray}
where we have used the condition $\tilde{g}(0) = 0$. Thus, due to the
condition $\tilde{g}(0) = 0$ the momentum integral $J(\lambda)$ does
not depend on $\lambda$
\begin{eqnarray}\label{label3.17}
\hspace{-0.5in}&&J(\lambda) = \int
\frac{d^2k}{2\pi}\,|\tilde{h}_{\lambda}(k)|^2\,F(k) =
\frac{\sigma}{2\pi}\,|\tilde{f}(0)|^2
\int^{\infty}_0\frac{dk_+}{k_+}\,|\tilde{g}(k_+)|^2,
\end{eqnarray}
The same independence can be obtained for $J_0(\lambda)$ and
$J_{00}(\lambda)$
\begin{eqnarray}\label{label3.18}
\hspace{-0.7in}&&J_0(\lambda) = \int
\frac{d^2k}{2\pi}\,|\tilde{h}_{\lambda}(k)|^2 F^{(+)}_0(k) =
\frac{\sigma}{4\pi}\,|\tilde{f}(0)|^2 \int^{\infty}_0dk_+
|\tilde{g}(k_+)|^2,\nonumber\\
\hspace{-0.7in}&&J_{00}(\lambda) = \int
\frac{d^2k}{2\pi}|\tilde{h}_{\lambda}(k)|^2 F^{(+)}_{00}(k) =
\frac{\sigma}{8\pi}\,|\tilde{f}(0)|^2
\int^{\infty}_0dk_+k_+\,|\tilde{g}(k_+)|^2.
\end{eqnarray}
The Cauchy--Schwarz inequality reads
\begin{eqnarray}\label{label3.19}
&&\hspace{-0.3in}\frac{\sigma}{2\pi}|\tilde{f}(0)|^2
\int^{\infty}_0\frac{dk_+}{k_+}|\tilde{g}(k_+)|^2
\frac{\sigma}{8\pi}|\tilde{f}(0)|^2 \int^{\infty}_0dk_+\,k_+
|\tilde{g}(k_+)|^2\ge \Bigg[\frac{\sigma}{4\pi}\,|\tilde{f}(0)|^2
\int^{\infty}_0dk_+|\tilde{g}(k_+)|^2\Bigg]^2\!\!.\nonumber\\
&&\hspace{-0.3in}
\end{eqnarray}
For $\tilde{f}(0) \neq 0$ and $\sigma \neq 0$ one can cancel the
common factor $(\sigma|\tilde{f}(0)|^2/4\pi)^2$ and get
\begin{eqnarray}\label{label3.20}
\int^{\infty}_0\frac{dk_+}{k_+}\,|\tilde{g}(k_+)|^2
\int^{\infty}_0dk_+\,k_+\,|\tilde{g}(k_+)|^2 \ge \Bigg[
\int^{\infty}_0dk_+\,|\tilde{g}(k_+)|^2\Bigg]^2.
\end{eqnarray}
Thus, it is seen that (i) on the sequence of the test functions
(\ref{label3.8}) from the Schwartz class ${\cal
S}(\mathbb{R}^{\,1})\otimes {\cal S}_0(\mathbb{R}^{\,1})$ the
Coleman's constraint is not a solution of the Cauchy--Schwarz
inequality (\ref{label3.7}) and (ii) setting $\sigma = 0$ the
Cauchy--Schwarz inequality (\ref{label3.19}) becomes a trivial
identity $0 \equiv 0$ but tells nothing about the non--existence of
Goldstone bosons.

This testifies that in Coleman's treatment massless (pseudo)scalar
quanta are excluded by the lemma (\ref{label3.10}) formulated on the
test functions from ${\cal S}(\mathbb{R}^{\,1})$. On the test
functions from ${\cal S}_0(\mathbb{R}^{\,1})$ the Cauchy--Schwarz
inequality (\ref{label3.19}) becomes a trivial identity $0\equiv 0$
due to $\tilde{f}(0) = 0$ for arbitrary $\sigma$. This confirms the
existence of the quantum field theory of a free massless
(pseudo)scalar field $\vartheta(x)$, described by the Lagrangian
(\ref{label1.2}), with Wightman's observables defined on the test
functions from ${\cal S}_0(\mathbb{R}^{\,2}) \supset {\cal
S}_0(\mathbb{R}^{\,1})\otimes {\cal S}_0(\mathbb{R}^{\,1})$. Hence, no
conclusion about the absence of spontaneously broken continuous
symmetry can be derived from Coleman's theorem with Coleman's lemma
formulated for the test functions $f(x_{\pm}) \in {\cal
S}_0(\mathbb{R}^{\,1})$.

\section{Quantum field theory of a massless (pseudo)scalar field 
in K\"allen--Lehmann representation for two--point Wightman functions}
\setcounter{equation}{0}

\hspace{0.2in} In this Section we construct a canonical quantum field
theory of a massless self--coupled (pseudo)scalar field which
satisfies Wightman's axioms and {\it Wightman's positive definiteness
condition} with Wightman's observables defined on the test functions
from ${\cal S}(\mathbb{R}^{\,2})$. In such a theory the symmetry,
related to the field--shifts (\ref{label1.3}), is spontaneously broken
and Goldstone bosons are the quanta of the massless (pseudo)scalar
field.

Let such a massless (pseudo)scalar field $\vartheta(x)$ be described
by the Lagrangian
\begin{eqnarray}\label{label4.1}
{\cal L}(x) = {\cal L}[\partial_{\mu}\vartheta(x)].
\end{eqnarray}
Due to the dependence on $\partial_{\mu}\vartheta(x)$ the Lagrangian
(\ref{label4.1}) is invariant under the field--shifts
(\ref{label1.3}). The current $j_{\mu}(x)$ related to the symmetry
transformation (\ref{label1.3}) is defined by
\begin{eqnarray}\label{label4.2}
j_{\mu}(x) = \frac{\delta {\cal L}[\partial_{\mu}\vartheta(x)]}{\delta
\partial^{\mu}\vartheta(x)}.
\end{eqnarray}
This current is conserved $\partial_{\mu}j^{\mu}(x) = 0$. The
conjugate momentum $\Pi(x)$ of the $\vartheta$--field is equal to the
time--component of the current
\begin{eqnarray}\label{label4.3}
\Pi(x)= \frac{\delta {\cal L}[\partial_{\mu}\vartheta(x)]}{\delta
\dot{\vartheta}(x)} = j_0(x),
\end{eqnarray}
where $\dot{\vartheta}(x)$ is a time derivative. The canonical
equal--time commutation relation (\ref{label1.6}) and the expression
of the $\vartheta$--field variation (\ref{label1.9}) are retained for
the quantum field theory of the massless $\vartheta$--field described
by the Lagrangian (\ref{label4.1}).

The most useful tool for the analysis of the Fourier transforms
(\ref{label3.1}) of the two--point correlation functions, defined in
this theory, is the K\"allen--Lehmann representation
\cite{HL54}. Inserting a complete set of intermediate states we
redefine the r.h.s. of the Fourier transforms (\ref{label3.1}) as
follows
\begin{eqnarray}\label{label4.4}
\hspace{-0.3in}&&F^{(+)}(k)= \int d^2x\,e^{\textstyle\,+ik\cdot
x}\,\sum_{n}\int d^2x\,e^{\textstyle + ik\cdot x}\,
\langle \Psi_0 |\vartheta(x)|n\rangle \langle n|\vartheta(0)|\Psi_0
\rangle =\nonumber\\ \hspace{-0.35in}&&= \int d^2x\,e^{\textstyle +
ik\cdot x}\langle\Psi_0 |\vartheta(x)|\Psi_0 \rangle\langle \Psi_0
|\vartheta(0) |\Psi_0 \rangle + \sum_{n\neq \Psi_0 }\int
d^2x\,e^{\textstyle + ik\cdot x}\langle \Psi_0 |\vartheta(x)|n\rangle
\langle n|\vartheta(0)|\Psi_0 \rangle,\nonumber\\
\hspace{-0.3in}&&F^{(+)}_{\mu}(k) = i\sum_{n}\int d^2x\,e^{\textstyle +
ik\cdot x}\langle \Psi_0 |j_{\mu}(x)|n\rangle\langle
n|\vartheta(0)|\Psi_0 \rangle,\nonumber\\
\hspace{-0.3in}&&F^{(+)}_{\mu\nu}(k) = \sum_{n}\int
d^2x\,e^{\textstyle + ik\cdot x}\langle
\Psi_0 |j_{\mu}(x)|n\rangle\langle n|j_{\nu}(0)|\Psi_0 \rangle.
\end{eqnarray}
Due to the invariance of the vacuum state $|\Psi_0 \rangle$ under space
and time translations and Lorentz covariance
$\langle\Psi_0 |j_{\mu}(x)|\Psi_0 \rangle = 0$, we have
\begin{eqnarray}\label{label4.5}
\hspace{-0.3in}F^{(+)}(k) &=& |\langle\Psi_0 |\vartheta(0)|\Psi_0
\rangle|^2 (2\pi)^2\,\delta^{(2)}(k) + \sum_{n\neq \Psi_0 }\int
d^2x\,e^{\textstyle + ik\cdot x}\langle \Psi_0 |\vartheta(x)|n\rangle
\langle n|\vartheta(0)|\Psi_0 \rangle,\nonumber\\
\hspace{-0.3in}F^{(+)}_{\mu}(k) &=& i\sum_{n\neq \Psi_0 } \int
d^2x\,e^{\textstyle + ik\cdot x}\langle
\Psi_0 |j_{\mu}(x)|n\rangle\langle
n|\vartheta(0)|\Psi_0 \rangle,\nonumber\\
\hspace{-0.3in}F^{(+)}_{\mu\nu}(k) &=& \sum_{n\neq \Psi_0 }\int
d^2x\,e^{\textstyle + ik\cdot x}\, \langle
\Psi_0 |j_{\mu}(x)|n\rangle\langle n|j_{\nu}(0)|\Psi_0 \rangle.
\end{eqnarray}
We would like to emphasize that the term proportional to
$\delta^{(2)}(k)$ in the Fourier transform of the Wightman function
appears only for a spontaneously broken symmetry and a non--invariant
vacuum because $\langle\Psi_0 |\vartheta(0)|\Psi_0 \rangle \neq 0$
\cite{EW78}.

Using again the invariance of the vacuum state $|\Psi_0 \rangle$ under
space and time translations we obtain \cite{HL54} 
\begin{eqnarray}\label{label4.6}
\hspace{-0.5in}F^{(+)}(k) &=& |\langle\Psi_0 |\vartheta(0)|\Psi_0
\rangle|^2\,(2\pi)^2\, \delta^{(2)}(k) + \theta(k^0)\int^{\infty}_0
\delta(k^2 - m^2)\,\rho_S(m^2)dm^2,\nonumber\\
\hspace{-0.5in}F^{(+)}_{\mu}(k) &=&-
\varepsilon_{\mu\nu}\,k^{\nu}\,\varepsilon(k^1)\,
\theta(k^0)\int^{\infty}_0 \delta(k^2 -
m^2)\,\rho_V(m^2)dm^2,\nonumber\\
\hspace{-0.5in}F^{(+)}_{\mu\nu}(k) &=&(k_{\mu}k_{\nu} - k^2
g_{\mu\nu})\theta(k^0)\int^{\infty}_0 \delta(k^2 -
m^2)\,\rho_T(m^2)dm^2,
\end{eqnarray}
where $\rho_i(m^2)\,(i = S, V, T)$ are the K\"allen--Lehmann spectral
functions defined by \cite{HL54}
\begin{eqnarray}\label{label4.7}
\hspace{-0.3in}&&(2\pi)^2 \sum_{n\neq \Psi_0 }\delta^{(2)}(k -
p_n)|\langle n|\vartheta(0)|\Psi_0 \rangle|^2 =
\theta(k^0)\int^{\infty}_0\!\!\!\delta(k^2 - m^2)\,\rho_S(m^2)dm^2
,\nonumber\\
\hspace{-0.3in}&&(2\pi)^2\sum_{n\neq \Psi_0 }\delta^{(2)}(k -
p_n)\langle \Psi_0 |j_{\mu}(0)|n\rangle\langle n|\vartheta(0)|\Psi_0
\rangle= \nonumber\\
\hspace{-0.3in}&&\hspace{1in}= i\varepsilon_{\mu\nu}\,k^{\nu}\,
\varepsilon(k^1)\,\theta(k^0) \int^{\infty}_0\!\!\!\delta(k^2 -
m^2)\,\rho_V(m^2)dm^2,\nonumber\\
\hspace{-0.3in}&&(2\pi)^2 \sum_{n\neq \Psi_0 }\delta^{(2)}(k -
p_n)\langle \Psi_0 |j_{\mu}(0)|n\rangle\langle n|j_{\nu}(0)|\Psi_0
\rangle=\nonumber\\
\hspace{-0.3in}&&\hspace{1in}= (k_{\mu}k_{\nu} - k^2
g_{\mu\nu})\theta(k^0)\int^{\infty}_0 \!\!\!\delta(k^2 -
m^2)\,\rho_T(m^2)dm^2.
\end{eqnarray}
We notice that for the massless state $- \varepsilon_{\mu\nu}
\,k^{\nu}\, \varepsilon(k^1) = k_{\mu}$.

Thus, in the K\"allen--Lehmann representation the Fourier transforms
$F^{(+)}(k)$, $F^{(+)}_{\mu}(k)$ and $F^{(+)}_{\mu\nu}(k)$ are defined
by
\begin{eqnarray}\label{label4.8}
F^{(+)}(k) &=& |\langle\Psi_0 |\vartheta(0)|\Psi_0 \rangle|^2 (2\pi)^2
\delta^{(2)}(k) + \theta(k^0)\int^{\infty}_0 \delta(k^2 - m^2)
\rho_S(m^2)dm^2,\nonumber\\
F^{(+)}_{\mu}(k) &=&- \varepsilon_{\mu\nu}\,k^{\nu}\,\varepsilon(k^1)\,
\theta(k^0)\int^{\infty}_0 \delta(k^2 -
m^2)\,\rho_V(m^2)dm^2,\nonumber\\
F^{(+)}_{\mu\nu}(k) &=&(k_{\mu}k_{\nu} - k^2
g_{\mu\nu})\theta(k^0)\int^{\infty}_0 \!\!\!\delta(k^2 -
m^2)\,\rho_T(m^2)dm^2.
\end{eqnarray}
These are the most general forms of distributions in the quantum field
theory of a massless self--coupled (pseudo)scalar field $\vartheta(x)$
in 1+1--dimensional space--time satisfying Wightman's axioms {\bf W1}
-- {\bf W4} and current conservation $\partial^{\mu}j_{\mu}(x) = 0$.

Now let us analyse {\it Wightman's positive definiteness condition} on
the test functions $h(x)$ from the Schwartz class ${\cal
S}(\mathbb{R}^{\,2})$. We get
\begin{eqnarray}\label{label4.9}
\hspace{-0.3in}&&\langle h, h\rangle = \int\!\!\!\int d^2x
d^2y\,h^*(x)\,D^{(+)}(x - y)\,h(y) = \int
\frac{d^2k}{(2\pi)^2}\,|\tilde{h}(k)|^2\,F^{(+)}(k) = \nonumber\\
\hspace{-0.3in}&&= |\langle\Psi_0 |\vartheta(0)|\Psi_0 \rangle|^2
|\tilde{h}(0)|^2 +
\int^{\infty}_0dm^2\,\rho_S(m^2)\int\frac{d^2k}{(2\pi)^2}\,
|\tilde{h}(k)|^2\,\theta(k^0)\,\delta(k^2 - m^2) =\nonumber\\
\hspace{-0.3in}&&= |\langle\Psi_0 |\vartheta(0)|\Psi_0 \rangle|^2
|\tilde{h}(0)|^2\nonumber\\
\hspace{-0.3in}&& + \int^{\infty}_0dm^2\,\rho_S(m^2)
\int^{\infty}_{-\infty}\int^{\infty}_{-\infty}
\frac{dk_+dk_-}{8\pi^2}\,|\tilde{h}(k_+,k_-)|^2\,\theta(k_+)\,
\theta(k_-)\,\delta(k_+k_- - m^2)=\nonumber\\
\hspace{-0.3in}&&= \langle\Psi_0 |\vartheta(0)|\Psi_0 \rangle^2
|\tilde{h}(0)|^2 + \frac{1}{8\pi^2}\int^{\infty}_0dm^2\,\rho_S(m^2)
\int^{\infty}_0\frac{dk_+}{k_+}\,
\Big|\tilde{h}\Big(k_+,\frac{m^2}{k_+}\Big)\Big|^2 \ge 0.
\end{eqnarray}
It is convenient to rewrite the second term as follows
\begin{eqnarray}\label{label4.10}
\hspace{-0.3in}&&\int^{\infty}_0\frac{dm^2}{8\pi^2}\,\rho_S(m^2)
\int^{\infty}_0\frac{dk_+}{k_+}\,
\Big|\tilde{h}\Big(k_+,\frac{m^2}{k_+}\Big)\Big|^2 =
\frac{\rho_S(0)}{8\pi^2}\int^{\infty}_0dk_+\int^{\infty}_0
\frac{dm^2}{k_+}
\Big|\tilde{h}\Big(k_+,\frac{m^2}{k_+}\Big)\Big|^2\nonumber\\
\hspace{-0.3in}&&+ \int^{\infty}_0\frac{dm^2}{8\pi^2}\,[\rho_S(m^2) -
\rho_S(0)]\,\int^{\infty}_0\frac{dk_+}{k_+}\,
\Big|\tilde{h}\Big(k_+,\frac{m^2}{k_+}\Big)\Big|^2
\end{eqnarray}
In the first integral we suggest to make a change of variables
$m^2/k_+ = k_-$. This gives
\begin{eqnarray}\label{label4.11}
\hspace{-0.3in}&&\int^{\infty}_0\frac{dm^2}{8\pi^2}\,\rho_S(m^2)
\int^{\infty}_0\frac{dk_+}{k_+}\,
\Big|\tilde{h}\Big(k_+,\frac{m^2}{k_+}\Big)\Big|^2 =
\rho_S(0)\int^{\infty}_0\int^{\infty}_0\frac{dk_+dk_-}{8\pi^2}
|\tilde{h}(k_+,k_-)|^2\nonumber\\
\hspace{-0.3in}&&+ \int^{\infty}_0\frac{dm^2}{8\pi^2}\,[\rho_S(m^2) -
\rho_S(0)]\,\int^{\infty}_0\frac{dk_+}{k_+}\,
\Big|\tilde{h}\Big(k_+,\frac{m^2}{k_+}\Big)\Big|^2.
\end{eqnarray}
Substituting (\ref{label4.11}) in (\ref{label4.9}) we obtain 
\begin{eqnarray}\label{label4.12}
\hspace{-0.3in}&&\langle h, h\rangle = \int\!\!\!\int d^2x
d^2y\,h^*(x)\,D^{(+)}(x - y)\,h(y) =\nonumber\\
\hspace{-0.3in}&&= |\langle\Psi_0 |\vartheta(0)|\Psi_0 \rangle|^2
|\tilde{h}(0)|^2 +
\rho_S(0)\int^{\infty}_0\int^{\infty}_0\frac{dk_+dk_-}{8\pi^2}
\,|\tilde{h}(k_+,k_-)|^2\nonumber\\
\hspace{-0.3in}&& + \int^{\infty}_0\frac{dm^2}{8\pi^2}\, [\rho_S(m^2)
- \rho_S(0)]\,\int^{\infty}_0\frac{dk_+}{k_+}\,
\Big|\tilde{h}\Big(k_+,\frac{m^2}{k_+}\Big)\Big|^2 \ge 0.
\end{eqnarray}
This testifies the fulfillment of the positive definiteness of the
scalar product $\langle h, h\rangle \ge 0$ on the test functions $h(x)
\in {\cal S}(\mathbb{R}^{\,2})$ in the quantum field theory of a
massless self--coupled (pseudo)scalar field $\vartheta(x)$ with the
Wightman functions defined by (\ref{label4.8}).

Our change of variable can be illustrated by an example.
\begin{eqnarray}\label{label4.13}
&&\int^{\infty}_0dm^2\int^{\infty}_0 \frac{dk_+}{k_+}
\Big|\tilde{h}\Big(k_+,\frac{m^2}{k_+}\Big)\Big|^2 =
\int^{\infty}_0dk_+\int^{\infty}_0 \frac{dm^2}{k_+}
\Big|\tilde{h}\Big(k_+,\frac{m^2}{k_+}\Big)\Big|^2 =\nonumber\\
&&=\int^{\infty}_0dm^2\int^{\infty}_0
\frac{dk_+}{k_+}\,\frac{1}{4}\,\exp\Big\{ - \frac{1}{2}\Big(k_+ +
\frac{m^2}{k_+}\Big)\Big\} = \int^{\infty}_0dm\,m\,K_0(m) = 1.
\end{eqnarray}
After the change of variable $m^2/k_+ = k_-$ we get
\begin{eqnarray}\label{label4.14}
&&\int^{\infty}_0dm^2\int^{\infty}_0
\frac{dk_+}{k_+}\frac{1}{4}\,\exp\Big\{ - \frac{1}{2}\Big(k_+ +
\frac{m^2}{k_+}\Big)\Big\} =\nonumber\\ &&=
\int^{\infty}_0dk_+\int^{\infty}_0dk_-\frac{1}{4}\,\exp\Big\{-
\frac{1}{2}(k_+ + k_-)\Big\} = 1.
\end{eqnarray}
It is easy to show that the Wightman function $D^{(+)}(x)$ as well as
the Fourier transform $F^{(+)}(k)$ is a tempered distribution defined
on the test functions $h(x)\in {\cal S}(\mathbb{R}^{\,2})$. For this
aim we have to calculate the functional $(h,D^{(+)})$ given by
\begin{eqnarray}\label{label4.15}
\hspace{-0.3in}&&(h^*,D^{(+)}) = \int d^2x\,h^*(x)\,D^{(+)}(x) = \int
\frac{d^2k}{(2\pi)^2}\,\tilde{h}^*(k)\,F^{(+)}(k) = |\langle\Psi_0
|\vartheta(0)|\Psi_0 \rangle|^2\tilde{h}^*(0)\nonumber\\
\hspace{-0.3in}&&+
\frac{1}{8\pi^2}\int^{\infty}_0dm^2\,\rho_S(m^2)\int^{\infty}_0
\frac{dk_+}{k_+}\,\tilde{h}^*\Big(k_+,\frac{m^2}{k_+}\Big) =
\langle\Psi_0 |\vartheta(0)|\Psi_0 \rangle^2\tilde{h}^*(0)+
\frac{\rho_S(0)}{8\pi^2}\int^{\infty}_0dm^2\nonumber\\
\hspace{-0.3in}&&\times\int^{\infty}_0
\frac{dk_+}{k_+}\,\tilde{h}^*\Big(k_+,\frac{m^2}{k_+}\Big) +
\frac{1}{8\pi^2}\int^{\infty}_0dm^2\,[\rho_S(m^2) -
\rho_S(0)]\int^{\infty}_0
\frac{dk_+}{k_+}\,\tilde{h}^*\Big(k_+,\frac{m^2}{k_+}\Big)
=\nonumber\\ \hspace{-0.3in}&&= \langle\Psi_0 |\vartheta(0)|\Psi_0
\rangle^2\tilde{h}^*(0) + \rho_S(0)\int^{\infty}_0\int^{\infty}_0
\frac{dk_+dk_-}{8\pi^2}\,\tilde{h}^*(k_+,k_-) \nonumber\\
\hspace{-0.3in}&&+ \frac{1}{8\pi^2}\int^{\infty}_0dm^2\,[\rho_S(m^2) -
\rho_S(0)]\int^{\infty}_0
\frac{dk_+}{k_+}\,\tilde{h}^*\Big(k_+,\frac{m^2}{k_+}\Big).
\end{eqnarray}
The r.h.s. of (\ref{label4.13}) contains only convergent
integrals. This testifies that the Wightman function
$D^{(+)}(x)$ and the Fourier transform $F^{(+)}(k)$ are
tempered distributions for Schwartz's test functions $h(x) \in {\cal
S}(\mathbb{R}^{\,2})$ satisfying {\it Wightman's positive definiteness
condition}.

Due to the necessity to fulfill {\it Wightman's positive definiteness
condition}, imposing to keep $\rho_S(0)$ finite or zero, we suggest
that the contribution of the state with $m^2 = 0$ is screened in the
Fourier transform $F^{(+)}(k)$. This is, of course, a dynamical effect
caused by the self--coupling of the $\vartheta$--field leading to the
influence of all intermediate states $|n\rangle$. Therefore, distinct
contributions of the state with $m^2 = 0$ can be only to the Fourier
transforms $F^{(+)}_{\mu}(k)$ and $F^{(+)}_{\mu\nu}(k)$. Isolating the
contributions of the state with $m^2 = 0$ in the spectral functions
$\rho_V(m^2)$ and $\rho_T(m^2)$ and setting
\begin{eqnarray}\label{label4.16}
\rho_V(m^2) &=& \sigma\,\delta(m^2) + \rho\,'_V(m^2),\nonumber\\
\rho_T(m^2) &=& \sigma '\delta(m^2) + \rho\,'_T(m^2),
\end{eqnarray}
we obtain the Fourier transforms $F^{(+)}(k)$, $F^{(+)}_{\mu}(k)$ and
$F^{(+)}_{\mu\nu}(k)$ in the following form
\begin{eqnarray}\label{label4.17}
\hspace{-0.5in}F^{(+)}(k) &=& |\langle\Psi_0 |\vartheta(0)|\Psi_0
\rangle|^2 (2\pi)^2 \delta^{(2)}(k) + \theta(k^0)\int^{\infty}_0
\delta(k^2 - m^2) \rho_S(m^2)dm^2,\nonumber\\
\hspace{-0.5in}F^{(+)}_{\mu}(k)
&=&\sigma k_{\mu} \theta(k^0) \delta(k^2) -
\varepsilon_{\mu\nu}\,k^{\nu}\,\varepsilon(k^1)\,
\theta(k^0)\int^{\infty}_{M^2} \delta(k^2 -
m^2) \rho\,'_V(m^2)dm^2,\nonumber\\
\hspace{-0.5in}F^{(+)}_{\mu\nu}(k) &=&\sigma' k_{\mu} k_{\nu}
\theta(k^0) \delta(k^2) + (k_{\mu}k_{\nu} - k^2
g_{\mu\nu})\theta(k^0)\int^{\infty}_{M^2} \!\!\!\delta(k^2 -
m^2)\rho\,'_T(m^2)dm^2,
\end{eqnarray}
where the spectral functions $\rho\,'_V(m^2)$ and $\rho\,'_T(m^2)$
contain only the contributions of the states with $m^2 > 0$ and the
scale $M^2$ isolates the state with $m^2 = 0$ from the states with
$m^2 > 0$.

The originals of the Fourier transforms given by (\ref{label4.8}) are
defined by
\begin{eqnarray}\label{label4.18}
D^{(+)}(x ) &=& \langle\Psi_0 |\vartheta(0)|\Psi_0 \rangle^2 +
\frac{1}{4\pi^2} \int^{\infty}_0dm^2\,\rho_S(m^2)\,K_0(m\sqrt{-x^2 +
i0\cdot\varepsilon(x^0)}),\nonumber\\ iD^{(+)}_{\mu}(x) &=& -i
\varepsilon_{\mu\nu}\frac{\partial}{\partial
x_{\nu}}\frac{1}{8\pi^2}\int^{\infty}_0dm^2\,\rho_V(m^2)
\int^{\varphi_0}_{-\varphi_0}d\varphi\,e^{\textstyle -m\sqrt{-x^2 +
i0\cdot \varepsilon(x^0)}\,{\cosh}\varphi},\nonumber\\
D^{(+)}_{\mu\nu}(x) &=&(\Box\,g_{\mu\nu} -
\partial_{\mu}\partial_{\nu})\frac{1}{4\pi^2}
\int^{\infty}_0dm^2\,\rho_T(m^2)\,K_0(m\sqrt{-x^2 +
i0\cdot\varepsilon(x^0)}),
\end{eqnarray}
where $K_0(m\sqrt{-x^2 + i0\cdot\varepsilon(x^0)})$ is the McDonald
function and $\varphi_0$ is defined by \cite{FI3}
\begin{eqnarray}\label{label4.19}
\varphi_0 = \frac{1}{2}\,{\ell n}\Big(\frac{x^0 + x^1 - i0}{x^0 - x^1
- i0}\Big).
\end{eqnarray}
In the original $iD^{(+)}_{\mu}(x)$ of the Fourier transform
$F^{(+)}_{\mu}(k)$ we suggest to isolate the contribution of the state
with $m^2 = 0$ from the contributions of the states with $m^2 >
0$. For this aim we transcribe the r.h.s. of $iD^{(+)}_{\mu}(x)$ as
follows
\begin{eqnarray}\label{label4.20}
&&iD^{(+)}_{\mu}(x) = -i \varepsilon_{\mu\nu}\frac{\partial
\varphi_0}{\partial
x_{\nu}}\Big[\frac{1}{4\pi^2}\int^{\infty}_0dm^2\,\rho_V(m^2)\Big]
\nonumber\\ && - i \varepsilon_{\mu\nu}\frac{\partial}{\partial
x_{\nu}}\frac{1}{8\pi^2}\int^{\infty}_0dm^2\,\rho_V(m^2)
\int^{\varphi_0}_{-\varphi_0}d\varphi\,\Big(e^{\textstyle -m\sqrt{-x^2
+ i0\cdot \varepsilon(x^0)}\,{\cosh}\varphi} - 1\Big)=\nonumber\\ &&=
\Big[\frac{1}{2\pi}\int^{\infty}_0 dm^2\,\rho_V(m^2)\Big]
\frac{i}{2\pi}\,\frac{x_{\mu}}{-x^2 + i0\cdot \varepsilon(x^0)}
\nonumber\\ && - i \varepsilon_{\mu\nu}\frac{\partial}{\partial
x_{\nu}}\frac{1}{8\pi^2}\int^{\infty}_0dm^2\,\rho_V(m^2)
\int^{\varphi_0}_{-\varphi_0}d\varphi\,\Big(e^{\textstyle -m\sqrt{-x^2
+ i0\cdot \varepsilon(x^0)}\,{\cosh}\varphi} - 1\Big)=\nonumber\\ &&=
\int^{\infty}_0dm^2\,\rho_V(m^2)\int
\frac{d^2q}{(2\pi)^2}\,q_{\mu}\,\theta(q^0)\,\delta(q^2)\,e^{\textstyle
-iq\cdot x}\nonumber\\ && - i
\varepsilon_{\mu\nu}\frac{\partial}{\partial
x_{\nu}}\frac{1}{8\pi^2}\int^{\infty}_0dm^2\,\rho_V(m^2)
\int^{\varphi_0}_{-\varphi_0}d\varphi\,\Big(e^{\textstyle -m\sqrt{-x^2
+ i0\cdot \varepsilon(x^0)}\,{\cosh}\varphi} - 1\Big).
\end{eqnarray}
This defines $iD^{(+)}_{\mu}(x)$ in the following general form
\begin{eqnarray}\label{label4.21}
\hspace{-0.3in}&&iD^{(+)}_{\mu}(x)
=\int^{\infty}_0dm^2\,\rho_V(m^2)\int
\frac{d^2q}{(2\pi)^2}\,q_{\mu}\,\theta(q^0)\,\delta(q^2)\,e^{\textstyle
-iq\cdot x}\nonumber\\ \hspace{-0.3in}&& - i
\varepsilon_{\mu\nu}\frac{\partial}{\partial
x_{\nu}}\frac{1}{8\pi^2}\int^{\infty}_0dm^2\,\rho_V(m^2)
\int^{\varphi_0}_{-\varphi_0}d\varphi\,\Big(e^{\textstyle -m\sqrt{-x^2
+ i0\cdot \varepsilon(x^0)}\,{\cosh}\varphi} - 1\Big).
\end{eqnarray}
The first term describes the contribution of the state with $m^2 = 0$,
whereas the second one contains the contributions of all states with
$m^2 > 0$.  Since the contribution of the state with $m^2 = 0$ is
defined by the expression $F^{(+)}_{\mu}(k; m^2 = 0)
=\sigma\,k_{\mu}\,\theta(k^0)\,\delta(k^2)$ we get the sum rules for
the spectral function $\rho_V(m^2)$, which read
\begin{eqnarray}\label{label4.22}
\int^{\infty}_0dm^2\,\rho_V(m^2) = \sigma.
\end{eqnarray}
Now let us consider the vacuum expectation value $\langle \Psi_0
|[j_{\mu}(x),\vartheta(0)]|\Psi_0 \rangle$. Following the standard
procedure expounded above we get
\begin{eqnarray}\label{label4.23}
\hspace{-0.5in}&&\langle
\Psi_0 |[j_{\mu}(x),\vartheta(0)]|\Psi_0 \rangle = \nonumber\\
\hspace{-0.5in}&& = i\int^{\infty}_0dm^2\,\rho_V(m^2)\int
\frac{d^2k}{(2\pi)^2}\,\varepsilon_{\mu\nu}\,k^{\nu}\,
\varepsilon(k^1)\,\theta(k^0)\,\delta(k^2 - m^2)\,(e^{\textstyle
-ik\cdot x} + e^{\textstyle +ik\cdot x}).
\end{eqnarray}
The vacuum expectation value of the the equal--time commutation
relation for the time--component of the current $j_0(0,x^1)$ and the
field $\vartheta(0)$ reads
\begin{eqnarray}\label{label4.24}
&&\langle \Psi_0 |[j_0(0,x^1),\vartheta(0)]|\Psi_0 \rangle =
-\frac{1}{2\pi}\int^{\infty}_0dm^2\,\rho_V(m^2)\,i\,\delta(x^1)
\nonumber\\&&- i\int^{\infty}_0dm^2\,\rho_V(m^2)\int^{\infty}_0
\frac{dk^1}{2\pi^2}\Big(\frac{k^1}{\sqrt{(k^1)^2 + m^2}} -
1\Big)\,\cos(k^1x^1).
\end{eqnarray}
It is seen that the state with $m^2 = 0$ does not contribute to the
second term. Therefore, the second term in the r.h.s. of
(\ref{label4.24}) is defined by only the spectral function
$\rho\,'_V(m^2)$. This gives
\begin{eqnarray}\label{label4.25}
&&\langle \Psi_0 |[j_0(0,x^1),\vartheta(0)]|\Psi_0 \rangle =
-\frac{1}{2\pi}\int^{\infty}_0dm^2\,\rho_V(m^2)\,i\,\delta(x^1)
\nonumber\\&&- i\int^{\infty}_{M^2}dm^2\,\rho\,'_V(m^2)\int^{\infty}_0
\frac{dk^1}{2\pi^2}\Big(\frac{k^1}{\sqrt{(k^1)^2 + m^2}} -
1\Big)\,\cos(k^1x^1).
\end{eqnarray}
Since time--component of the current $j_{\mu}(x)$ coincides with the
conjugate momentum, i.e. $\Pi(x) = j_0(x)$, the l.h.s. of
(\ref{label4.25}) is equal to $-i\,\delta(x^1)$. Using the canonical
equal--time commutation relation (\ref{label1.6}) for the l.h.s. of
(\ref{label4.24}) we derive the sum rules
\begin{eqnarray}\label{label4.26}
\int^{\infty}_0dm^2\,\rho_V(m^2) = 2\pi.
\end{eqnarray}
Comparing (\ref{label4.26}) with (\ref{label4.22}) we get
\begin{eqnarray}\label{label4.27}
\sigma = 2\pi.
\end{eqnarray}
This is a model--independent result which rules out Coleman's result
asserting $\sigma = 0$.  

As the second term in the r.h.s. of (\ref{label4.25}) can be never
proportional to $\delta(x^1)$ it should be identically zero. This
yields
\begin{eqnarray}\label{label4.28}
\rho\,'_V(m^2) \equiv 0.
\end{eqnarray}
Hence, the spectral function $\rho_V(m^2)$ is equal to
\begin{eqnarray}\label{label4.29}
\rho_V(m^2) = \sigma\,\delta(m^2) = 2\pi\,\delta(m^2).
\end{eqnarray}
This means that in the case of current conservation
$\partial^{\mu}j_{\mu}(x) = 0$ the Fourier transform
$F^{(+)}_{\mu}(k)$ is defined by only the contribution of the state
with $m^2 = 0$. This confirms that the expression (\ref{label2.4})
postulated by Coleman is general for the {\it canonical} quantum field
theories with conserved current $\partial^{\mu}j_{\mu}(x) = 0$ in
1+1--dimensional quantum field theories of a massless (pseudo)scalar
field $\vartheta(x)$, but this rules out Coleman's constraint $\sigma
= 0$. 

We would like to emphasize that in the {\it non--canonical} quantum
field theory, whether such a theory could exist, for which $j_0(x)
\neq \Pi(x)$, the parameter $\sigma$ can be arbitrary but the spectral
function $\rho\,'_V(m^2)$ does not vanish. This means that if Coleman
would have considered a non--canonical quantum field theory, the
equation (13) of Ref.\cite{SC73} for the Fourier transform
$F^{(+)}_{\mu}(k)$ should contain this term. Recall that the second
term in (13) of Ref.\cite{SC73} is caused by parity violation and
is not related to the contribution of $\rho\,'_V(m^2)$, which
conserves parity. Dropping the term caused by $\rho\,'_V(m^2)$ Coleman
has asserted implicitly that he is in the framework of a canonical
quantum field theory.

Multiplying (\ref{label4.24}) by $i \alpha$ and integrating over $x^1$
we obtain $\langle \Psi_0 |\delta \vartheta(0)|\Psi_0 \rangle$, which
reads
\begin{eqnarray}\label{label4.30}
\langle \Psi_0 |\delta \vartheta(0)|\Psi_0  \rangle =
i\,\alpha\int^{\infty}_{-\infty}dx^1\,\langle
\Psi_0 |[j_{\mu}(x),\vartheta(0)]|\Psi_0 \rangle = \frac{\alpha}{2\pi}
\int^{\infty}_0dm^2\,\rho_V(m^2) = \alpha,
\end{eqnarray}
where we have taken into account the discussion above and the
expression for the spectral function $\rho_V(m^2)$ given by
(\ref{label4.29}). In agreement with the Goldstone theorem \cite{JG61}
the non--vanishing value of the $\vartheta$--field variation
(\ref{label4.30}) testifies the existence of Goldstone bosons and
spontaneously broken continuous symmetry (\ref{label1.3}).

In order to analyse the properties of the spectral function
$\rho_T(m^2)$ we suggest to calculate the vacuum expectation value of
the commutator $[j_{\mu}(x),j_{\nu}(0)]$. In terms of the spectral
function $\rho_T(m^2)$ the result reads
\begin{eqnarray}\label{label4.31}
\hspace{-0.5in}&&\langle \Psi_0 |[j_{\mu}(x),j_{\nu}(0)]|\Psi_0 \rangle
=\nonumber\\
\hspace{-0.5in}&&= \int^{\infty}_0dm^2 \rho_T(m^2)\int
\frac{d^2k}{(2\pi)^2}\,(k_{\mu}k_{\nu} -
k^2g_{\mu\nu})\theta(k^0)\delta(k^2 - m^2)\,(e^{\textstyle -ik\cdot x}
- e^{\textstyle +ik\cdot x}).
\end{eqnarray}
Considering the component $\langle \Psi_0 |[j_0(x),j_1(0)]|\Psi_0
\rangle$ at $x^0 = 0$ we get
\begin{eqnarray}\label{label4.32}
\hspace{-0.5in}&&\langle \Psi_0 |[j_0(0,x^1),j_1(0)]|\Psi_0 \rangle
=\nonumber\\
\hspace{-0.5in}&&= \int^{\infty}_0dm^2 \rho_T(m^2)\int
\frac{d^2k}{(2\pi)^2}\,k_0k_1\theta(k^0)\delta(k^2 - m^2)\,
(e^{\textstyle +ik^1x^1} - e^{\textstyle -ik^1x^1})=\nonumber\\
\hspace{-0.5in}&&= \frac{1}{2\pi}\int^{\infty}_0dm^2
\rho_T(m^2)\,i\frac{\partial}{\partial x^1}\delta(x^1).
\end{eqnarray}
Since due to Schwinger \cite{JS59} the commutator $[j_0(0,x^1),
j_1(0)]$ is defined by
\begin{eqnarray}\label{label4.33}
[j_0(0,x^1), j_1(0)] = i S \frac{\partial}{\partial x^1}\delta(x^1),
\end{eqnarray}
where $S$ is a Schwinger term, we get the sum rules
\begin{eqnarray}\label{label4.34}
\int^{\infty}_0dm^2 \rho_T(m^2) = 2\pi S.
\end{eqnarray}
This testifies that the integral over $m^2$ of the spectral function
$\rho_T(m^2)$ is convergent.

The analysis of Wightman's observables $\vartheta(h)$ on the test
functions $h(x)$ from ${\cal S}(\mathbb{R}^{\,2})$ and ${\cal
S}_0(\mathbb{R}^{\,2})$, which we have carried out in Section 2, is
applicable to the canonical quantum field theory of a massless
self--coupled (pseudo)scalar field $\vartheta(x)$ formulated in this
Section. Indeed, for the general case $h(x) \in {\cal
S}(\mathbb{R}^{\,2})$ the Wightman observable $\vartheta(h)$ is not
invariant under the field--shifts (\ref{label1.3}). We get 
\begin{eqnarray}\label{label4.35}
\vartheta\,'(h) = e^{\textstyle +i\alpha
Q(x^0)}\vartheta(h)e^{\textstyle -i\alpha Q(x^0)} = \vartheta(h) +
\alpha\int d^2x\,h(x).
\end{eqnarray}
This yields the variation of Wightman's observable 
\begin{eqnarray}\label{label4.36}
\delta \vartheta(h) = \alpha\int d^2x\,h(x).
\end{eqnarray}
It is important to emphasize that $\delta \vartheta(h)$ given by
(\ref{label4.36}) is not an operator--valued quantity. Due to this the
vacuum expectation value coincides with the quantity
\begin{eqnarray}\label{label4.37}
\langle \Psi_0|\delta \vartheta(h)|\Psi_0\rangle = \delta \vartheta(h)
=\alpha\int d^2x\,h(x).
\end{eqnarray}
Hence, the variation of Wightman observable $\delta \vartheta(h)$ can
say nothing about spontaneous breaking of continuous symmetry. 

Then, narrowing  the class of the test functions from ${\cal
S}(\mathbb{R}^{\,2})$ to ${\cal S}_0(\mathbb{R}^{\,2})$ one gets the
Wightman observable $\vartheta(h)$ invariant under shifts
(\ref{label1.3}) and the variation of Wightman's observable $\delta
\vartheta(h)$ identically zero, $\delta \vartheta(h) = 0$. However,
this does not give a new information about the Goldstone bosons and a
spontaneously broken continuous symmetry in addition to that we have
got on the class of the test functions from ${\cal
S}(\mathbb{R}^{\,2})$.

The necessity to use the test functions from ${\cal
S}_0(\mathbb{R}^{\,2})$ for the definition of Wightman's observables
in the quantum field theory of a self--coupled massless (pseudo)scalar
field, described by the Lagrangian (\ref{label4.1}), can be
demonstrated by example of the definition of the generating functional
of Green functions. Denoting the generating functional of Green
functions as ${\cal Z}[J]$ we define \cite{NB59}
\begin{eqnarray}\label{label4.38}
{\cal Z}[J] &=& \Big\langle \Psi_0\Big|{\rm T}\Big(e^{\textstyle i\int
d^2x\,\vartheta(x)J(x)}\Big|\Psi_0\Big\rangle = \nonumber\\ &=&\int
{\cal D}\vartheta\,e^{\textstyle i\int d^2x\,\{{\cal
L}[\partial_{\mu}\vartheta(x)] +
\vartheta(x)J(x)\}}\Big|\Psi_0\Big\rangle = \nonumber\\
&=&e^{\textstyle i\int d^2x\,{\cal L}\,'\Big( -
i\partial_{\mu}\frac{\textstyle \delta }{\textstyle \delta
J(x)}\Big)}Z[J],
\end{eqnarray}
where $Z[J]$ is determined by (\ref{label1.17}) and ${\cal L}\,'$ is
the Lagrangian of the self--interaction of the massless (pseudo)scalar
field $\vartheta(x)$
\begin{eqnarray}\label{label4.39}
{\cal L}\,'[\partial_{\mu}\vartheta(x)] = {\cal
L}[\partial_{\mu}\vartheta(x)] -
\frac{1}{2}\,\partial_{\mu}\vartheta(x)\partial^{\mu}\vartheta(x).
\end{eqnarray}
Since in order to get a non--vanishing value for $Z[J]$ one needs the
constraint (\ref{label1.18}), the collective zero--mode is deleted
from the intermediate states of correlation functions. This result
agrees well with Hasenfratz \cite{PH84} who showed that the removal of
the collective zero--mode motion of the self--coupled massless scalar
fields in lattice $\sigma$--models with $O(N)$ symmetry in one and two
dimensional volumes does not affect the evolution of the system but
allows to construct a self--consistent perturbation theory for the
calculation of correlation functions.

This means that in the quantum field theory of a self--coupled
massless (pseudo)scalar field $\vartheta(x)$, described by the
Lagrangian (\ref{label4.1}), Wightman's observables should be defined
on the test functions $h(x)$ from ${\cal S}_0(\mathbb{R}^{\,2})$.

\section{Conclusion}
\setcounter{equation}{0}

\hspace{0.2in} We have shown that Coleman's constraint $\sigma = 0$,
interpreted as a proof for the absence of Goldstone bosons and
spontaneously broken continuous symmetry in 1+1--dimensional quantum
field theories, is not a consequence of the Cauchy--Schwarz
inequality, but follows from Coleman's lemma. Formulating the lemma on
the test functions from ${\cal S}(\mathbb{R}^{\,1})\otimes {\cal
S}_0(\mathbb{R}^{\,1})$ Coleman has followed Wightman's axioms
demanding the definition of Wightman's observables on the test
functions from ${\cal S}(\mathbb{R}^{\,2}) \supset {\cal
S}(\mathbb{R}^{\,1})\otimes {\cal S}_0(\mathbb{R}^{\,1})$. Due to his
lemma Coleman removed the massless state from the Fourier transform of
the Wightman function, that has entailed the exclusion of this state
from all two--point correlation functions. This agrees with Wightman's
assertion about the non--existence of the quantum field theory of a
free massless (pseudo)scalar field in 1+1--dimensional space--time,
when Wightman's observables are defined on the test functions $h(x)$
from the Schwartz class ${\cal S}(\mathbb{R}^2)$, $h(x) \in {\cal
S}(\mathbb{R}^2)$.

However, we have motivated the reduction of the test functions $h(x)$
from the Schwartz class ${\cal S}(\mathbb{R}^{\,2})$ to ${\cal
S}_0(\mathbb{R}^{\,2}) = \{h(x) \in {\cal S}(\mathbb{R}^{\,2});
\tilde{h}(0) = 0\}$. The definition of Wightman's observables on the
test functions from ${\cal S}_0(\mathbb{R}^{\,2}) = \{h(x) \in {\cal
S}(\mathbb{R}^{\,2});\tilde{h}(0) = 0\}$ instead of ${\cal
S}(\mathbb{R}^{\,2})$ is justified by the removal of the collective
zero--mode from the massless (pseudo)scalar field $\vartheta(x)$.  The
collective zero--mode does not affect the evolution of a free massless
(pseudo)scalar field. 

Wightman's observables defined on the test functions from ${\cal
S}_0(\mathbb{R}^{\,2})$ do not measure the collective zero--mode due
to the constraint $\tilde{h}(0) = 0$. This makes a quantum field
theory of a free massless (pseudo)scalar field well--defined within
Wightman's axiomatic approach.

The removal of the collective zero--mode agrees well with Hasenfratz's
approach to one and two dimensional quantum field theories
\cite{PH84}, who showed that a self--consistent perturbation theory
for the calculation of correlation functions in $\sigma$--models for
self--coupled massless scalar fields with $O(N)$ symmetry can be
developed only removing the collective zero--mode.

The reformulation of Coleman's lemma for the test functions from
${\cal S}_0(\mathbb{R}^{\,1})$ reduces the Cauchy--Schwarz inequality
to the identity $0 \equiv 0$ for an arbitrary parameter $\sigma$. This
certifies that Coleman's analysis of 1+1--dimensional quantum field
theories does not contradict the existence of Goldstone bosons and
spontaneous breaking of continuous symmetry in the quantum field
theory of the free and the self--coupled massless (pseudo)scalar field
for Wightman's observables defined on the test functions from the
Schwartz class ${\cal S}_0(\mathbb{R}^2)$.

The physical states of quanta of vibrational modes of a free massless
(pseudo)scalar field in a quantum field theory with Wightman's
observables, defined on the test functions from ${\cal
S}_0(\mathbb{R}^2)$, are determined in a {\it positive--definite}
Hilbert space \cite{AW64}. According to Nakanishi \cite{NN03} the use
of test functions from the Schwartz class ${\cal S}_0(\mathbb{R}^2)$
leads to a violation of {\it localizability} of the Schwartz
distributions. This entails the impossibility to define consistently
the {\it support} of the Schwartz distributions \cite{NN03,LS57}. Due
to this the formulation of a quantum field theory of a free massless
(pseudo)scalar field in 1+1--dimensional space--time with Wightman's
observables, defined on the test functions from the Schwartz class
${\cal S}(\mathbb{R}^2)$, and physical states, determined in an {\it
indefinite--metric} Hilbert space, is more advantageous \cite{NN03}. 

An analysis of Coleman's theorem for a two--dimensional quantum field
theory of a free massless (pseudo)scalar field with Wightman's
observables, defined on the test functions from the Schwartz class
${\cal S}(\mathbb{R}^2)$, and physical states, determined in an {\it
indefinite--metric} Hilbert space, we are planning to perform in a
forthcoming publication.

An interesting analysis of Coleman's theorem \cite{SC73}, applied to
the 1+1--dimensional quantum field theory of self--coupled ``charged''
bosons with an internal non--Abelian continuous symmetry, has been
recently suggested by Chigak Itoi \cite{Itoi03}.

\end{document}